\documentclass[journal]{IEEEtran}
\usepackage[noadjust]{cite}
\usepackage{srcltx}
\usepackage{graphicx}
\usepackage{epstopdf}
\usepackage{psfrag}
\usepackage{epsfig}
\usepackage[tbtags,sumlimits,nointlimits,reqno]{amsmath}
\usepackage{amssymb}
\usepackage{xcolor}
\usepackage{float}
\usepackage{subfigure}
\usepackage[most]{tcolorbox}
\usepackage{soul}
\usepackage{footnote}
\usepackage{amsmath}
\usepackage{color}
\newcommand{\jstyle}[1]{\textit{\textbf{{1}}}}
\newcommand{\rstyle}[1]{\textit{\textbf{{1}}}}
\newcommand{\cstyle}[1]{\textit{\textbf{{1}}}}
\newcommand{\bstyle}[1]{\textit{\textbf{{1}}}}
\newcommand{\istyle}[1]{\emph{\textbf{{1}}}}
\newcommand{\tbpstyle}[1]{\textit{{1}}}
\usepackage[framemethod=TikZ]{mdframed}
\usepackage{lipsum}
\mdfdefinestyle{MyFrame}{%
	linecolor=blue,
	outerlinewidth=2pt,
	roundcorner=20pt,
	innertopmargin=\baselineskip,
	innerbottommargin=\baselineskip,
	innerrightmargin=20pt,
	innerleftmargin=20pt,
	backgroundcolor=gray!50!white}

\begin{document}
	\title{Temporal Bragg Gratings:\\ Broadband Reconfigurable Parametric Amplifiers}
	\author{Sajjad Taravati,~\IEEEmembership{Senior Member,~IEEE}
		\thanks{S. Taravati is with the Faculty of Engineering and Physical Sciences, University of Southampton, Southampton SO17 1BJ, UK (e-mail: s.taravati@soton.ac.uk).}%
	\thanks{Manuscript received *, 2025; revised *, 2026.}
}
	\markboth{IEEE Transaction on Antennas and Propagation,~Vol.~*, No.~*, *~2026}%
	{* \MakeLowercase{\textit{et al.}}: Bare Demo of IEEEtran.cls for IEEE Journals}

	\maketitle
	
\begin{abstract}
This paper introduces temporal Bragg gratings as a new class of broadband, reconfigurable parametric amplifiers. We present a comprehensive investigation of power amplification in temporal Bragg gratings, spatially periodic structures with refractive index modulated near the Bragg frequency. Through systematic numerical simulations, we explore the effects of modulation location (high-index vs. low-index layers), frequency, and amplitude on gain spectra and field dynamics. Both layer types yield significant parametric amplification, with high-index modulation providing higher gain for comparable depths. Amplification is frequency-agile, with gain peaks tunable across a broad range, and exhibits strong asymmetry: the sub-Bragg regime (\(\omega_\text{m} < \omega_\text{B}\)) requires substantially stronger modulation than supra-Bragg for comparable gain. In the extreme sub-Bragg limit (\(\omega_\text{m} \lesssim 0.1\omega_\text{B}\)), the system transitions from discrete sidebands to a broadband gain continuum via multi-phase-matching. These results establish a unified framework for designing reconfigurable optical amplifiers, tunable frequency converters, and broadband light sources using temporally modulated photonic crystals.
\end{abstract}

	
	\IEEEpeerreviewmaketitle

\section{Introduction}\label{sec:introduction}
\IEEEPARstart{T}he control of light propagation through engineered materials has been revolutionized by the concepts of photonic crystals, gratings and metamaterials, which enable unprecedented manipulation of electromagnetic waves in both space and frequency domains~\cite{othonos1997fiber,hill2002fiber,ansari2020ultra,Taravati_NC_2021,taravati2024spatiotemporal,taravati20234d,taravati2024efficient,sisler2024electrically,taravati2025designing}. By introducing periodic variations in the refractive index, spatial photonic crystals and Bragg gratings, researchers have demonstrated remarkable phenomena such as photonic bandgaps, slow light, and enhanced nonlinear interactions for sensing and biomedical applications~\cite{burla2013integrated,albert2013tilted,rohan2024recent,la2024bragg}. More recently, the paradigm of temporal photonics has emerged, where material properties are modulated in time and space, unlocking fundamentally new degrees of freedom for controlling light~\cite{eichler2013laser,Taravati_Kishk_PRB_2018,Taravati_PRAp_2018,Taravati_Kishk_MicMag_2019,taravati_PRApp_2019,Taravati_Kishk_TAP_2019,taravati2020full,koufidis2023temporal,Taravati_ACSP_2022,taravati2024finite,taravati2024one,Taravati_AMTech_2021,alex2025kapitza,taravati2025light,wu2025space}.

Temporal modulation breaks time-reversal symmetry and enables nonreciprocal behavior~\cite{Taravati_PRB_SB_2017,Taravati_AMA_PRApp_2020,taravati2020full,das2025electrical,taravati2025designing}, frequency conversion~\cite{taravati2021pure,Taravati_PRB_Mixer_2018,moreno2024space,nadi2024beam,taravati2024efficient,wu2024synthetic,taravati2025frequency}, amplification without traditional gain media~\cite{cullen1958travelling,tien1958traveling,tien1958parametric,Taravati_Kishk_PRB_2018,li2019nonreciprocal,Taravati_AMA_PRApp_2020,sumetsky2025transformation}, and quantum frequency multiplexing and nonreciprocity~\cite{taravati2024spatiotemporal,taravati2025_entangle,taravati2025light,taravati2025frequency}. When combined with spatial periodicity, as in spatio-temporal photonic crystals or temporal Bragg gratings, rich interference phenomena arise that can lead to momentum bandgaps, parametric amplification, and frequency generation.

This study introduces temporal Bragg gratings, spatially periodic structures with coherent time modulation, and demonstrates their ability to provide huge, dynamically reconfigurable power amplification. Through comprehensive numerical and analytical analysis, we show how this platform transforms from a passive filter into an active parametric device. Systematic investigation of modulation location (high- vs. low-index layers), frequency, and amplitude reveals unprecedented control over gain strength and spectral agility. Our results reveal that frequency-agile amplification is achievable by detuning \(\omega_\text{m}\) from \(\omega_\text{B}\) which enables tunable gain peaks at frequencies \(\omega_\text{B} \pm \omega_\text{m}\). A pronounced asymmetry exists between the sub-Bragg (\(\omega_\text{m} < \omega_\text{B}\)) and supra-Bragg (\(\omega_\text{m} > \omega_\text{B}\)) regimes, with the former requiring significantly stronger modulation for comparable gain. In the extreme sub-Bragg limit (\(\omega_\text{m} \lesssim 0.1\omega_\text{B}\)), the system transitions from discrete sideband amplification to broadband gain at high frequencies, explained by multi-phase-matching of non-degenerate four-wave mixing processes. Both high-index and low-index modulation can produce strong amplification. Our findings establish a unified framework for amplification in temporal Bragg gratings, and positions them as versatile active photonic platforms with dynamically controllable spectral response.

\section{Theoretical Implications}\label{sec:theory}
\subsection{Temporal Bragg Grating Architecture}
\begin{figure}
	\begin{center}
		\includegraphics[width=1\columnwidth]{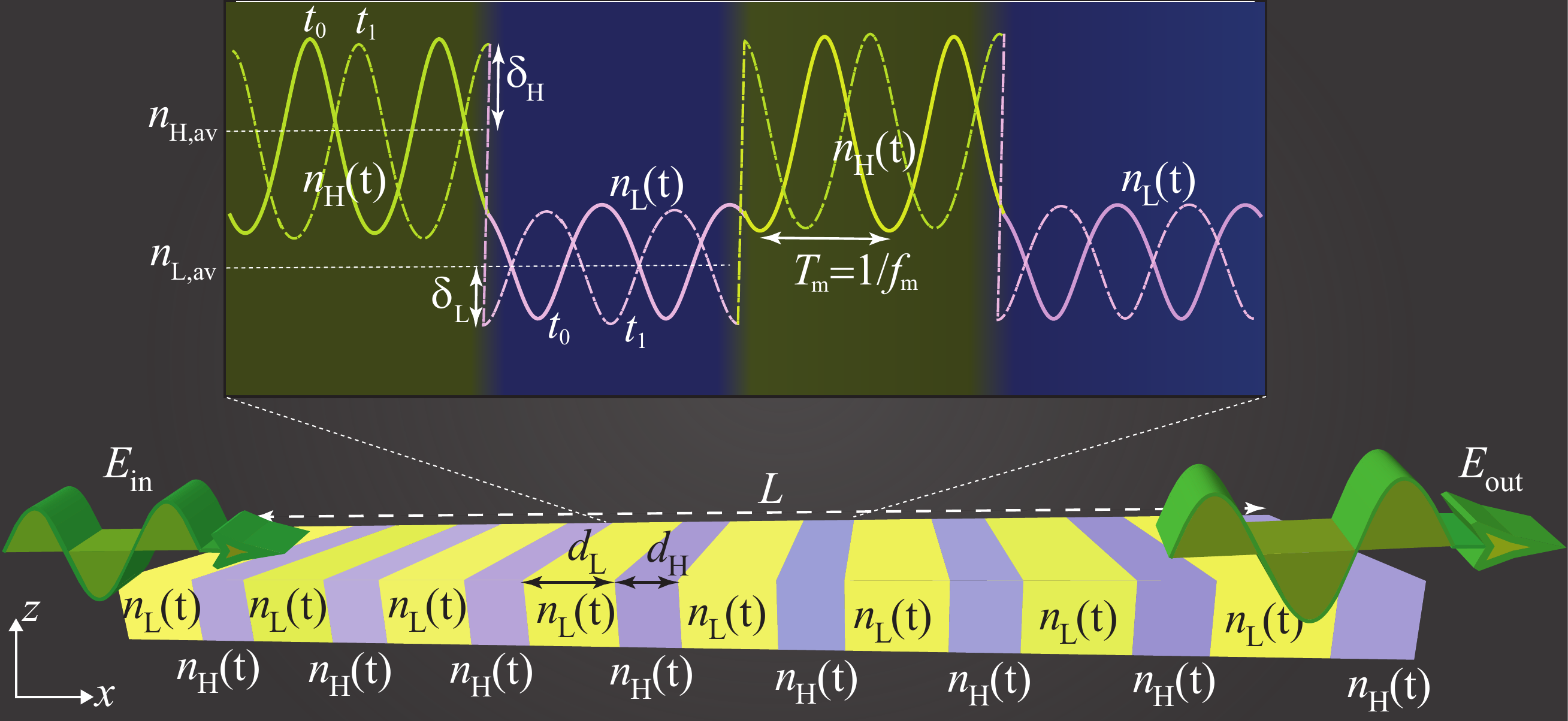}
		\caption{Schematic of a temporal Bragg grating for through-port parametric amplification, composing alternating time-periodic high-index (\( n_\text{H}(t) \)) and low-index (\( n_\text{L}(t) \)) refractive indices.\vspace{-0.5cm}}
		\label{Fig:sch}
	\end{center}
\end{figure}

A Bragg grating is a periodic structure composed of alternating layers with high (\(n_\text{H}\)) and low (\(n_\text{L}\)) refractive indices, each having an optical thickness of \(\lambda_\text{B}/4\) at the design wavelength \(\lambda_\text{B}\). This quarter-wave condition ensures constructive interference of reflections from each interface, creating a photonic stopband centered at the Bragg frequency \(\omega_\text{B} = 2\pi c/\lambda_\text{B}\). The temporal Bragg grating in Fig.~\ref{Fig:sch} assumes that either the high-index or low-index layers experience sinusoidal temporal modulation, as
\begin{equation}\label{eq:refractive_index_time}
	n(x,t) = 
\begin{cases}
			n_\text{H}(t) = n_{\text{H},\text{av}} + \Delta_\text{H} \cos(\omega_\text{m} t + \phi_\text{H}) \\
			n_\text{L}(t) = n_{\text{L},\text{av}} + \Delta_\text{L} \cos(\omega_\text{m} t + \phi_\text{L}),
\end{cases}
\end{equation}
where $\Delta_\text{H}=\delta_\text{H} n_\text{H}$ and $\Delta_\text{L}=\delta_\text{L}n_\text{L}$, in which $\delta_\text{H}$ and $\delta_\text{L}$ are the temporal modulation amplitudes of the high and low index layers, respectively. In addition, \(\omega_\text{m} = 2\pi f_\text{m}\) is the modulation frequency, \(\delta_{H}, \delta_{L} \ll n_{\text{av}}\) are the modulation depths, and \(\phi_\text{H}, \phi_\text{L}\) are modulation phases. The temporal Bragg grating consists of \(N\) periods, each with spatial period \(\Lambda\). The total physical length of the structure is $L = N \Lambda = N(d_\text{H} + d_\text{L})$, where \(d_\text{H} = \lambda_\text{B}/(4n_\text{H})\) and \(d_\text{L} = \lambda_\text{B}/(4n_\text{L})\) are the thicknesses of the high- and low-index layers, respectively. This length \(L\) determines the interaction distance for parametric processes and appears in all phase-matching conditions and gain expressions. This temporal variation breaks time-translation symmetry, enabling energy transfer from the modulation pump at frequency \(\omega_\text{m}\) to the optical signal at \(\omega_0\) via parametric photon conversion $\hbar\omega_{\text{out}} = \hbar\omega_0 \pm \hbar\omega_\text{m}$.

\subsection{Interference of Multi-Harmonic Multi-Scattering}\label{sec:multi_harmonic_scattering}

\begin{figure}
	\begin{center}
		\includegraphics[width=0.9\columnwidth]{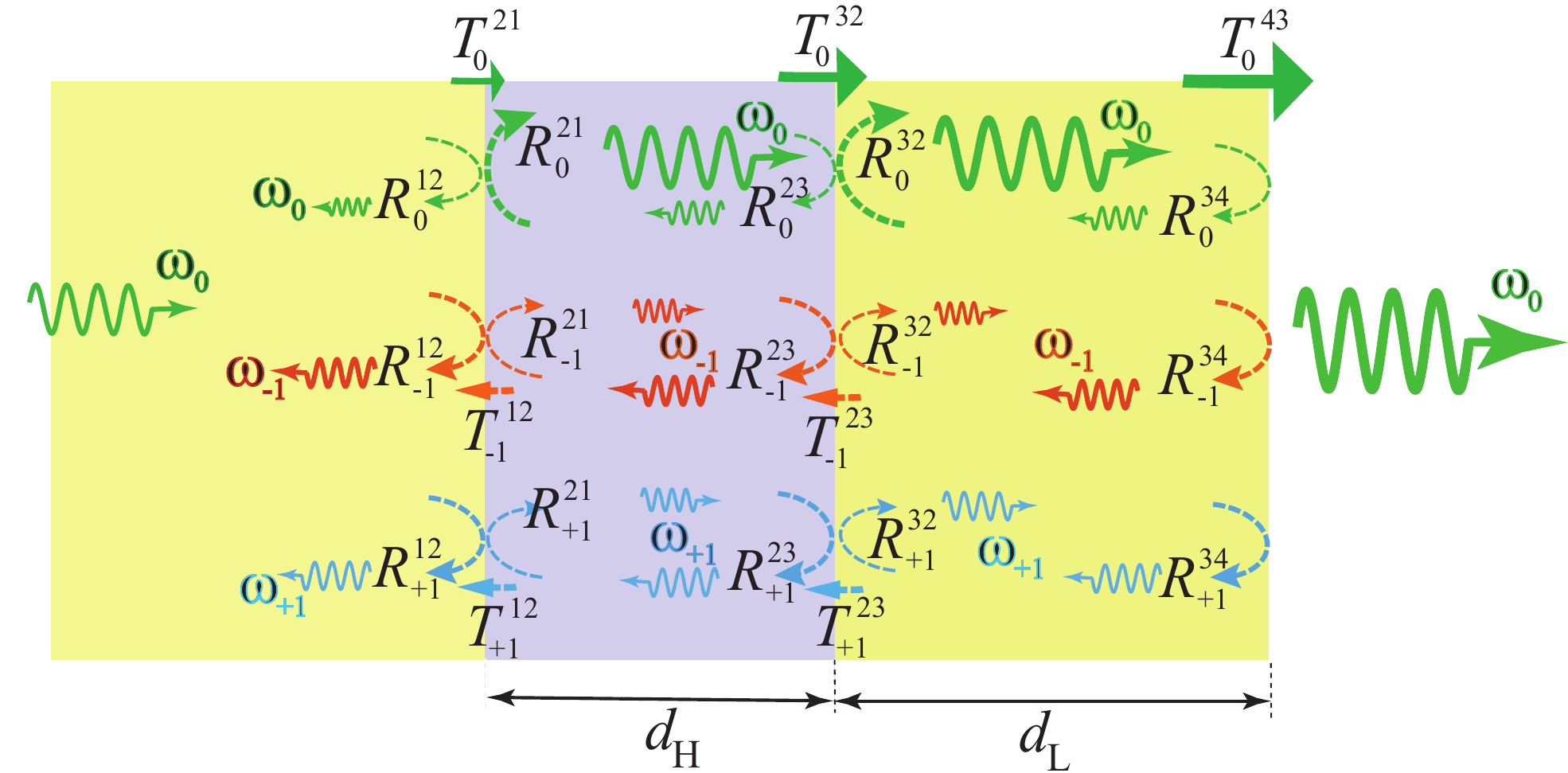}
		\caption{Amplification of \( \omega_0 \) arises from multiple reflections and transmissions of the main frequency \( \omega_0 \) and generated sidebands \( \omega_{\pm1} = \omega_0 \pm \omega_\text{m} \) through time-modulated layers. While \( \omega_{\pm1} \) are generated via parametric coupling at each interface, only \( \omega_0 \) experiences constructive interference across the structure, leading to net gain in transmission.\vspace{-0.5cm}}
		\label{Fig:multi}
	\end{center}
\end{figure}

The amplification mechanism in temporal Bragg gratings can be visualized as a distributed multi-harmonic scattering process, where the input wave at frequency $\omega_0$ interacts with time-modulated interfaces to generate sidebands at $\omega_{\pm1} = \omega_0 \pm \omega_\text{m}$, as illustrated in Fig.~\ref{Fig:multi}. These harmonics undergo multiple reflections and transmissions throughout the structure, with constructive interference selectively enhancing the fundamental frequency $\omega_0$. Consider a temporal Bragg grating composed of $N$ alternating layers with time-varying refractive indices $n_j(t) = n_{j,0} + \delta n_j \cos(\omega_\text{m} t + \phi_j)$. At each interface $j$ between layers $j$ and $j+1$, the time-varying Fresnel coefficients couple different temporal harmonics. For a monochromatic wave at frequency $\omega_p = \omega_0 + p\omega_\text{m}$, the standard scattering matrix at interface $j$ reads
\begin{equation}\label{eq:static_scattering_matrix}
	\mathbf{S}^{(j)}_p = \begin{pmatrix}
		r_p^{(j)} & t_p^{(j)} \\
		t_p^{(j)} & -r_p^{(j)}
	\end{pmatrix},
\end{equation}
where the Fresnel coefficients are:
\begin{equation}\label{eq:fresnel_coeffs}
	r_p^{(j)} = \frac{n_j(\omega_p) - n_{j+1}(\omega_p)}{n_j(\omega_p) + n_{j+1}(\omega_p)}, \quad
	t_p^{(j)} = \frac{2n_j(\omega_p)}{n_j(\omega_p) + n_{j+1}(\omega_p)}.
\end{equation}

Due to temporal modulation, the refractive indices become time-dependent, introducing coupling between harmonics. Expanding the time-varying Fresnel coefficients to first order in $\delta n_j$
\begin{equation}\label{eq:time_var_coeffs}
	\begin{split}
		r^{(j)}(t) &= r_0^{(j)} + \frac{\delta n_j}{2} \left(\frac{\partial r^{(j)}}{\partial n_j} e^{i\omega_\text{m} t} + \frac{\partial r^{(j)}}{\partial n_{j+1}} e^{i\omega_\text{m} t}\right) + \text{c.c.}, \\
		t^{(j)}(t) &= t_0^{(j)} + \frac{\delta n_j}{2} \left(\frac{\partial t^{(j)}}{\partial n_j} e^{i\omega_\text{m} t} + \frac{\partial t^{(j)}}{\partial n_{j+1}} e^{i\omega_\text{m} t}\right) + \text{c.c.}.
	\end{split}
\end{equation}

This temporal variation generates the block scattering matrix $\underline{\mathbf{S}}^{(j)}$ that couples harmonics $p$ and $q$, where the diagonal blocks $\mathbf{S}^{(j)}_{p,p}$ are given by Eq.~(\ref{eq:static_scattering_matrix}) and the off-diagonal coupling blocks are
\begin{equation}\label{eq:coupling_blocks}
	\mathbf{S}^{(j)}_{p,q} = \frac{\delta n_j}{2n_0} \begin{pmatrix}
		\Delta r^{(j)}_{pq} & \Delta t^{(j)}_{pq} \\
		\Delta t^{(j)}_{pq} & -\Delta r^{(j)}_{pq}
	\end{pmatrix}, \quad |p-q| = 1,
\end{equation}
where the coupling coefficients $\Delta r^{(j)}_{pq}$ and $\Delta t^{(j)}_{pq}$ derive from the derivatives in Eq.~(\ref{eq:time_var_coeffs}). For a grating with $N$ interfaces, the total scattering matrix is obtained by cascading individual matrices using the Redheffer star product $\star$, $\underline{\mathbf{S}}_{\text{total}} = \underline{\mathbf{S}}^{(1)} \star \underline{\mathbf{S}}^{(2)} \star \cdots \star \underline{\mathbf{S}}^{(N)}$. Define the harmonic amplitude vectors for forward ($\mathbf{F}^{(j)}$) and backward ($\mathbf{B}^{(j)}$) waves in layer $j$, i.e., $\mathbf{F}^{(j)} = [\dots, F_{-1}^{(j)}, F_0^{(j)}, F_{+1}^{(j)}, \dots]^T$, $\mathbf{B}^{(j)} = [\dots, B_{-1}^{(j)}, B_0^{(j)}, B_{+1}^{(j)}, \dots]^T$. The scattering relation for the entire structure reads
\begin{equation}\label{eq:total_scattering_relation}
	\begin{pmatrix}
		\mathbf{B}^{(1)} \\
		\mathbf{F}^{(N+1)}
	\end{pmatrix}
	= \underline{\mathbf{S}}_{\text{total}}
	\begin{pmatrix}
		\mathbf{F}^{(1)} \\
		\mathbf{B}^{(N+1)}
	\end{pmatrix}.
\end{equation}

From the scattering picture emerges a set of coupled amplitude equations that explicitly track the multiple reflection paths shown in Fig.~\ref{Fig:multi}. For harmonic $p$ in layer $j$:
\begin{subequations}\label{eq:coupled_amplitudes}
	\begin{align}
		A_p^{(j)} &= t_p^{(j-1)} A_p^{(j-1)} + r_p^{(j)} A_p^{(j+1)} \nonumber \\
		&\quad + \sum_{q=p\pm1} \left[ \kappa_{pq}^{(j)} A_q^{(j-1)} + \kappa_{pq}^{(j)\prime} A_q^{(j+1)} \right], \label{eq:amp_main} \\
		\kappa_{pq}^{(j)} &= \frac{\delta n_j}{2n_0} \frac{\partial t^{(j)}}{\partial n} e^{i(\phi_j + \beta_q d_j)}, 
		\label{eq:coupling_coeff}
	\end{align}
\end{subequations}
where $\beta_q = n_{\text{eff},q}\omega_q/c$ is the propagation constant for harmonic $q$, and $d_j$ is the thickness of layer $j$. The different interference conditions for $\omega_0$ versus $\omega_{\pm1}$ explain the observed selective amplification.

\textit{Amplification of $\omega_0$}
The multiple reflection paths for $\omega_0$ add coherently when the round-trip phase accumulation satisfies
\begin{equation}\label{eq:constructive_condition}
	2\sum_{j=1}^{N} \beta_0^{(j)} d_j + \arg\left(r_0^{(j)} r_0^{(j+1)}\right) = 2\pi m, \quad m \in \mathbb{Z},
\end{equation}
and parametric contributions from $\omega_{\pm1}$ via coupling terms $\kappa_{0,\pm1}$ arrive in phase, leading to constructive interference and exponential growth.

\textit{Suppression of $\omega_{\pm1}$}: The generated sidebands originate at different interfaces with phases $\phi_{\pm1}^{(j)} = \phi_j \pm \omega_\text{m} t_j$, where $t_j$ is the arrival time at interface $j$. These phase variations prevent coherent buildup, resulting in
\begin{equation}\label{eq:incoherent_sum}
	A_{\pm1}^{\text{total}} = \sum_{j=1}^{N} \kappa_{\pm1,0}^{(j)} A_0^{(j)} e^{i\phi_{\pm1}^{(j)}} \approx 0 \quad \text{(incoherent sum)}.
\end{equation}

The amplification can be understood as an effective Fabry-Perot resonance with parametrically enhanced reflectivity. Summing the geometric series of all reflection paths yields
\begin{equation}\label{eq:effective_reflection}
	R_{\text{eff}} = R_0 + \frac{|\kappa_{0,+1}|^2}{1 - R_{+1}} + \frac{|\kappa_{0,-1}|^2}{1 - R_{-1}},
\end{equation}
where $R_p$ is the reflection coefficient for harmonic $p$. The total power gain for $\omega_0$ is then
\begin{equation}\label{eq:total_gain_scattering}
	G_{\text{total}} = \left|\frac{t_0^{\text{(in)}} t_0^{\text{(out)}}}{1 - R_{\text{eff}} e^{i\Phi}}\right|^2,
\end{equation}
with $\Phi = 2\beta_0 L$ being the round-trip phase.

\subsection{Electromagnetic Fields}
Consider a temporal Bragg grating with refractive index modulation $n(\mathbf{r},t) = n_0(\mathbf{r}) + \delta n(\mathbf{r}) \cos(\omega_\text{m} t + \phi(\mathbf{r}))$, where $\delta n \ll n_0$, $\omega_\text{m}$ is the modulation frequency, and the modulation is applied selectively to either high-index ($n_H$) or low-index ($n_L$) layers. For a non-magnetic, lossless dielectric medium with time-varying refractive index, Maxwell's equations read
\begin{subequations}\label{eq:maxwell_time_var}
	\begin{equation}
		\nabla \times \mathbf{E} = -\mu_0 \frac{\partial \mathbf{H}}{\partial t} ;\quad 
		\nabla \times \mathbf{H} = \epsilon_0 \frac{\partial}{\partial t} \left[ n^2(\mathbf{r},t) \mathbf{E} \right].
		 \label{eq:maxwell2}
	\end{equation}

Considering a $z$-polarized wave propagating in the $x$-direction in a one-dimensional structure, combining Maxwell's equations, for small modulation depth ($\delta n \ll n_0$), yields
\begin{equation}\label{eq:wave_eq_time_var}
	\frac{\partial^2 E_z}{\partial x^2} - \frac{n^2(x,t)}{c^2} \frac{\partial^2 E_z}{\partial t^2} - \frac{1}{c^2} \frac{\partial n^2}{\partial t} \frac{\partial E_z}{\partial t} = 0.
\end{equation}
\end{subequations}

The structure has dual periodicity: spatial period $\Lambda$ (Bragg period) and temporal period $T_\text{m} = 2\pi/\omega_\text{m}$. According to Floquet-Bloch theorem, solutions take the form
\begin{subequations}
\begin{equation}\label{eq:floquet_bloch}
	E_z(x,t) = e^{j(\beta x - \omega_0 t)} \sum_{p=-\infty}^{\infty} \sum_{q=-\infty}^{\infty} A_{pq} e^{j(pKx + q\omega_\text{m} t)},
\end{equation}
where $\beta$ is the Bloch wavevector, $K = 2\pi/\Lambda$ is the grating wavevector, and $A_{pq}$ are the amplitudes of spatial harmonic $p$ and temporal harmonic $q$. For practical analysis considering only temporal harmonics (setting spatial harmonics $p=0$ for the fundamental mode), the field expands as
\begin{equation}\label{eq:field_expansion}
	E_z(x,t) = \sum_{p=-\infty}^{\infty} \left[ A_p^+(x) e^{j(\beta_p x - \omega_p t)} + A_p^-(x) e^{j(-\beta_p x - \omega_p t)} \right],
\end{equation}
\end{subequations}
where $\omega_p = \omega_0 + p\omega_\text{m}$, $\beta_p = n_{\text{eff}}(\omega_p)\omega_p/c$, and $n_{\text{eff}}$ is the effective index of the guided mode. Applying the slowly varying envelope approximation to Maxwell's equations yields coupled-mode equations. For forward-propagating waves:
\begin{subequations}
\begin{equation}\label{eq:coupled_mode_forward}
	\begin{split}
		j\frac{dA_p^+}{dx} = \kappa_p A_p^- e^{-j2\Delta_p x} +\frac{\omega_p}{c}\delta n(x) \sum_{q \neq p} C_{pq} A_q^+ e^{-j\Delta k_{pq}x},
	\end{split}
\end{equation}
and for backward-propagating waves
\begin{equation}\label{eq:coupled_mode_backward}
	\begin{split}
		-j\frac{dA_p^-}{dx} = \kappa_p A_p^+ e^{+j2\Delta_p x} + \frac{\omega_p}{c}\delta n(x) \sum_{q \neq p} C_{pq} A_q^- e^{-j\Delta k_{pq}x},
	\end{split}
\end{equation}
where $\kappa_p$ is the spatial coupling coefficient due to Bragg periodicity, $\Delta_p = \beta_p - \pi/\Lambda$ is detuning from the Bragg condition, and
\begin{equation}\label{eq:phase_mismatch}
	\Delta k_{pq} = \beta_p - \beta_q - (p-q)K,
\end{equation}
is the phase mismatch between harmonics $p$ and $q$. The coupling coefficients are determined by overlap integrals
\begin{equation}\label{eq:overlap_integral}
	C_{pq} = \frac{\int_{\text{modulated layers}} E_p^*(x) E_q(x) dx}{\sqrt{\int_{\text{structure}} |E_p(x)|^2 dx \int_{\text{structure}} |E_q(x)|^2 dx}}.
\end{equation}
\end{subequations}

\subsection{Parametric Amplification Mechanism}
For parametric amplification, we consider three interacting waves: pump ($f_\text{m}$), signal ($f_\text{s}$), and idler ($f_\text{i}$), with energy conservation $f_\text{s} + f_\text{i} = f_\text{m}$ (for non-degenerate case) or $f_\text{s} = f_\text{i} = f_\text{m}/2$ (for degenerate case). The coupled equations simplify to
	\begin{equation}\label{eq:three_wave_coupled}
		\frac{dA_\text{s}}{dx} = -j\gamma A_i^* e^{-j\Delta k_{\text{si}}x};\quad
		\frac{dA_i}{dx} = -j\gamma A_\text{s}^* e^{-j\Delta k_{\text{si}}x},
	\end{equation}
where $\gamma = \frac{\omega_\text{s} \delta n}{2c} \eta_{\text{overlap}}$ is the nonlinear coupling coefficient, $\eta_{\text{overlap}}$ is the spatial overlap factor from Eq.~\eqref{eq:overlap_integral}, and $\Delta k_{\text{si}} = \beta_\text{s} + \beta_\text{i} - \beta_\text{m} - K$ is the phase mismatch. When phase-matched ($\Delta k_{\text{si}} = 0$), the solution exhibits exponential growth
\begin{subequations}\label{eq:parametric_growth}
	\begin{align}
		A_\text{s}(x) &= A_\text{s}(0) \cosh(gx) + jA_\text{i}^*(0) \sinh(gx), \label{eq:signal_solution} \\
		A_\text{i}(x) &= A_\text{i}(0) \cosh(gx) + jA_\text{s}^*(0) \sinh(gx), \label{eq:idler_solution}
	\end{align}
\end{subequations}
with parametric gain coefficient
\begin{subequations}
\begin{equation}\label{eq:gain_coefficient}
	g = |\gamma| = \left|\frac{\omega_\text{s} \delta n}{2c} \cdot \eta_{\text{overlap}}\right|.
\end{equation}

The power gain in decibels after propagation through a structure of length $L$ is
\begin{equation}\label{eq:power_gain_dB}
	\begin{split}
	G_{\text{dB}} &= 10\log\left[\cosh^2(gL)\right] \approx 
	\begin{cases}
		4.343 (gL)^2 & \text{for } gL \ll 1 \\
		8.686 gL - 6.020 & \text{for } gL \gg 1.
	\end{cases}
		\end{split}
\end{equation}
\end{subequations}

\textit{Layer-Dependent Asymmetric Gain}: The spatial overlap factor $\eta_{\text{overlap}}$ differs significantly depending on whether modulation is applied to high-index or low-index layers. For high-index layer modulation
\begin{subequations}
\begin{equation}\label{eq:overlap_H}
	\eta_{\text{overlap}}^H \approx \frac{\int_{\text{high-index layers}} |E(x)|^2 dx}{\int_{\text{all layers}} |E(x)|^2 dx} > \frac{1}{2},
\end{equation}
since the field concentrates in high-index layers in a quarter-wave stack. For low-index layer modulation:
\begin{equation}\label{eq:overlap_L}
	\eta_{\text{overlap}}^L \approx \frac{\int_{\text{low-index layers}} |E(x)|^2 dx}{\int_{\text{all layers}} |E(x)|^2 dx} < \frac{1}{2}.
\end{equation}

This asymmetry in field overlap leads to different gain spectra: high-index modulation preferentially amplifies frequencies above $f_0$, while low-index modulation amplifies frequencies below $f_0$. The gain asymmetry ratio reads
\begin{equation}\label{eq:gain_asymmetry_ratio}
	\frac{G(f_0 + \Delta f)}{G(f_0 - \Delta f)} \approx \left(\frac{\eta_{\text{overlap}}^H(f_0 + \Delta f)}{\eta_{\text{overlap}}^L(f_0 - \Delta f)}\right)^2.
\end{equation}
\end{subequations}
	
\subsection{Resonant Modulation Conditions}
When the modulation frequency equals the Bragg frequency ($f_\text{m} = f_\text{B}$), the system operates in degenerate mode with $f_\text{s} = f_\text{i} = f_\text{B}/2$. The gain is enhanced by the cavity quality factor $Q$
\begin{equation}\label{eq:degenerate_gain}
	g_{\text{deg}} = \frac{\omega_0 \delta n}{4c} \cdot \eta_{\text{overlap}} \cdot Q,
\end{equation}
where $Q = f_\text{B}/\Delta f_{\text{FWHM}}$ characterizes the Bragg resonance linewidth. For $f_\text{m} = 2f_\text{B}$, the signal and idler are at $f_\text{B}$, creating a second-harmonic-type process. The phase-matching condition is given by
\begin{equation}\label{eq:phase_match_2w}
	\Delta k = 2\beta(f_\text{B}) - \beta(2f_\text{B}) - K = 0.
\end{equation}

This condition can be satisfied near the photonic band edge where dispersion is strong. The maximum gain occurs near $f_\text{m} = 1.5f_\text{B}$. This corresponds to signal and idler frequencies $f_\text{s} = 0.5f_\text{m} = 0.75f_\text{B}$ and $f_i = 0.5f_\text{m} = 0.75f_\text{B}$, placing both near the band edge where group velocity is minimal and interaction time is maximized.

\section{Results}
\subsection{Regime I: Coherent and Supra-Bragg Modulation}
\begin{figure}
	\begin{center}
		\valign{#\cr
			\hsize=0.5\columnwidth
			\subfigure[]{\label{Fig:Res1d}
				\includegraphics[width=.26\textwidth,height=3.2cm]{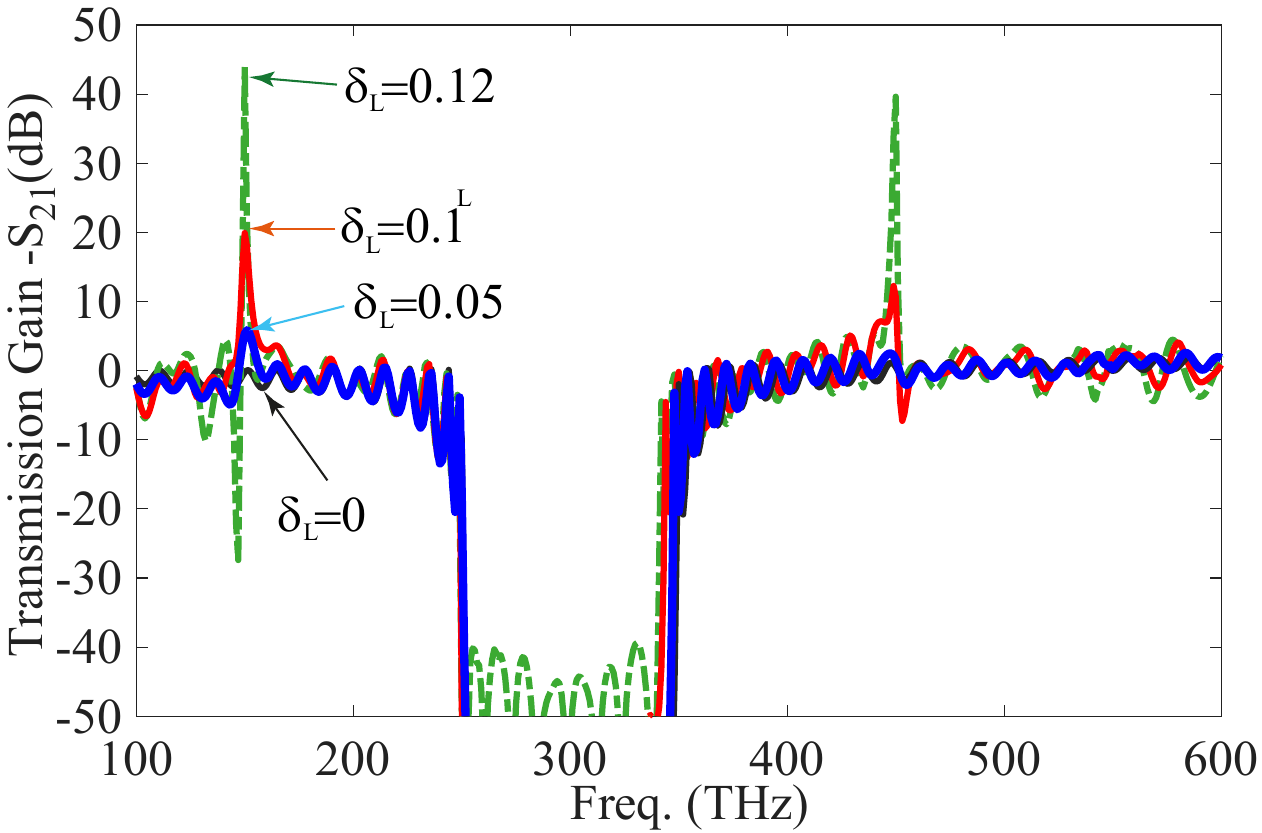}}\cr\noalign{\hfill}
			\hsize=0.45\columnwidth
			\subfigure[]{\label{Fig:Res1a}
				\includegraphics[width=0.425\columnwidth]{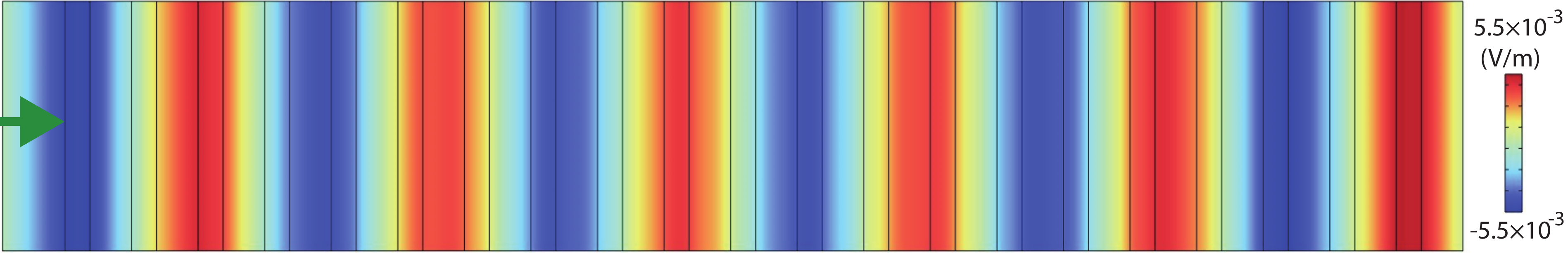}}
			\subfigure[]{\label{Fig:Res1b}
				\includegraphics[width=0.425\columnwidth]{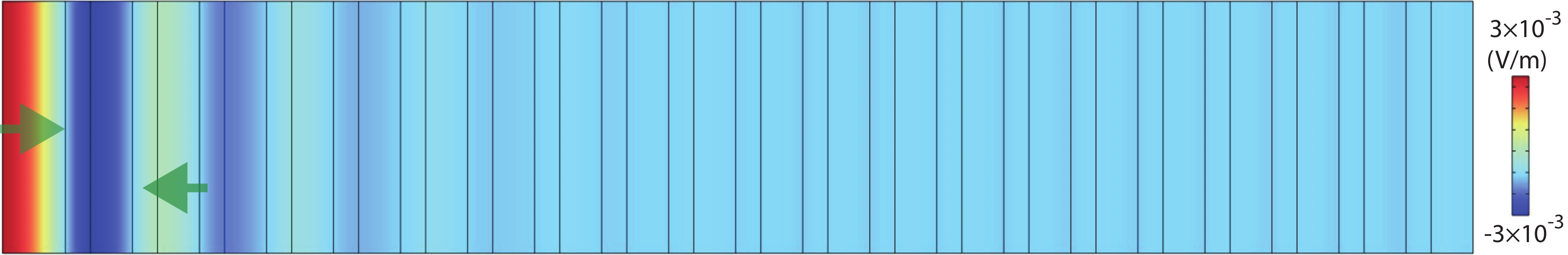}}
			\subfigure[]{\label{Fig:Res1c}
				\includegraphics[width=0.425\columnwidth]{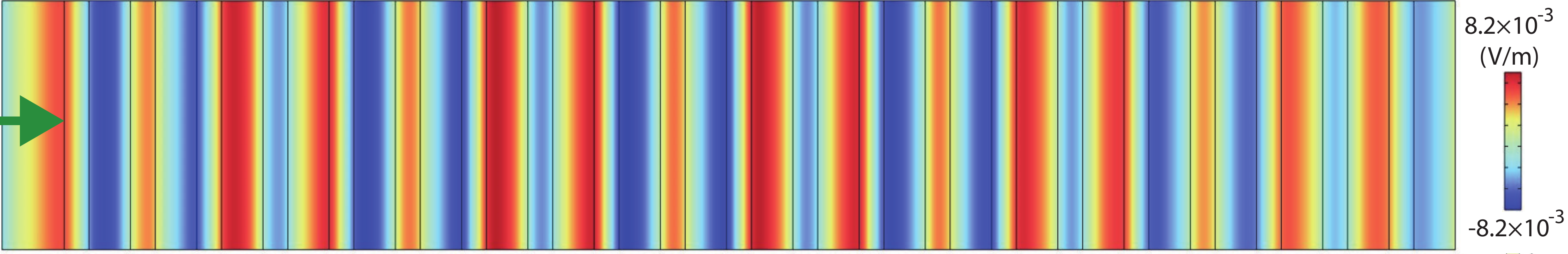}}\cr}\vspace{-0.5cm}
		\caption{Power amplification in a coherent temporal Bragg grating with a modulation frequency equal to the Bragg frequency, \(f_{\rm m} = f_{\rm B} = 300\) THz. The structure parameters are \(n_{\rm L}=1.5\), \(n_{\rm H}=2.5\), \(\lambda_{0}=1\ \mu\text{m}\), \(\Lambda=20\), \(\delta_{\rm H}=0\), and \(\delta_{\rm L}=0.12\). (a)~Transmission gain for different time-periodic low-index modulations \(\delta_{\rm L}>0\), with static high-index layers (\(\delta_{\rm H}=0\)). (b)-(d)~Normal electric field distribution at $f_{\rm B}/2$, $f_{\rm B}$, and $1.5f_{\rm B}$, respectively.\vspace{-0.5cm}}
		\label{Fig:Res1}
	\end{center}
\end{figure}

Figure~\ref{Fig:Res1} illustrates power amplification in a coherent temporal Bragg grating when the modulation frequency \(f_{\text{m}}\) is equal to the Bragg frequency \(f_{\text{B}} = 300\)~THz, corresponding to a central wavelength $\lambda_\text{B} = 1 \mu$m. The transmission spectrum (Fig.~\ref{Fig:Res1d}) further confirms that when the high-index layers are static (\(\delta_{\text{H}} = 0\)) and the low-index layers are time-modulated (\(\delta_{\text{L}} > 0\)), substantial gain occurs at frequencies symmetrically displaced from \(f_{\text{B}}\), notably at \(f_{\text{B}}/2\) and \(1.5f_{\text{B}}\), while transmission is suppressed at the Bragg frequency itself. The electric field distributions reveal distinct spectral behaviors. At \(f_{\text{B}}/2\), the field pattern shows clear amplification (Fig.~\ref{Fig:Res1a}). At \(f_{\text{B}}\), strong reflection is observed with no transmission, as expected at the main Bragg frequency (Fig.~\ref{Fig:Res1b}). At \(1.5f_{\text{B}}\), significant field amplification reappears (Fig.~\ref{Fig:Res1c}).

\begin{figure}
\begin{center}
	\subfigure[]{\label{Fig:Res11a}
		\includegraphics[width=0.48\columnwidth]{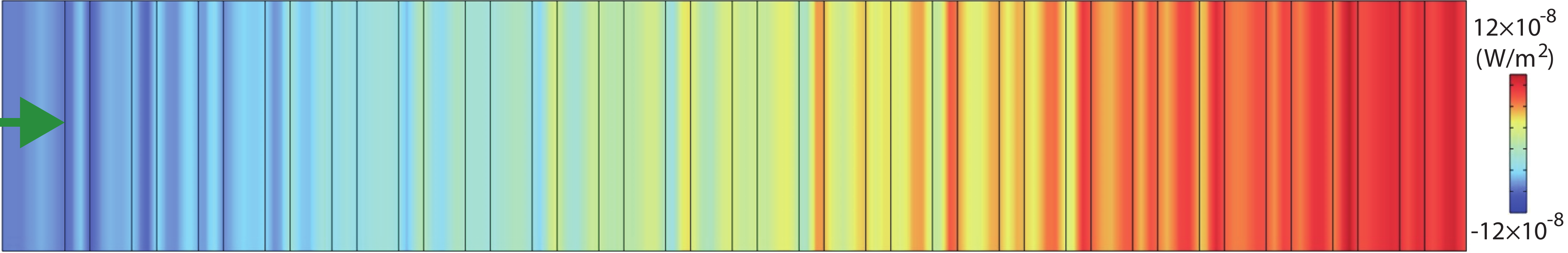}}
	\subfigure[]{\label{Fig:Res11b}
		\includegraphics[width=0.48\columnwidth]{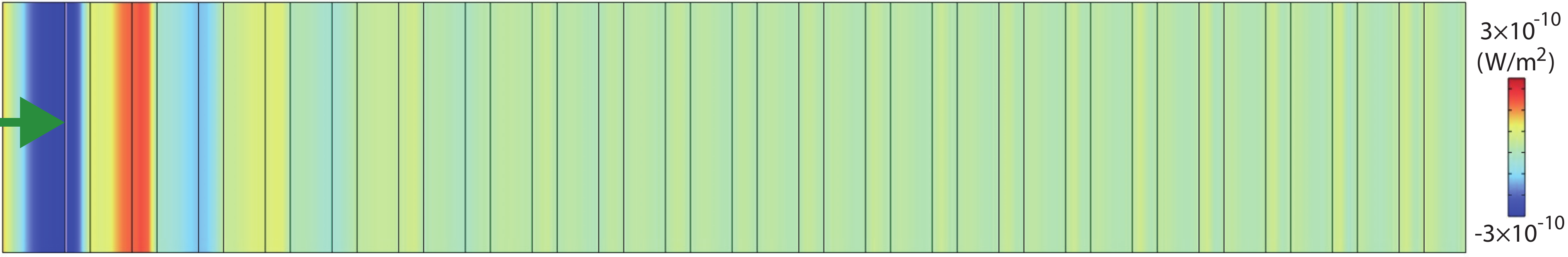}}
	\subfigure[]{\label{Fig:Res11c}
		\includegraphics[width=0.48\columnwidth]{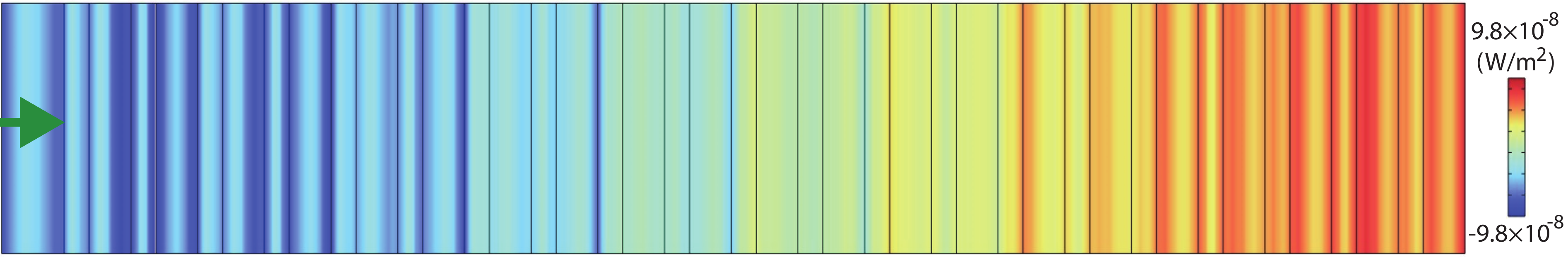}}
	\subfigure[]{\label{Fig:Res11d}
		\includegraphics[width=0.48\columnwidth]{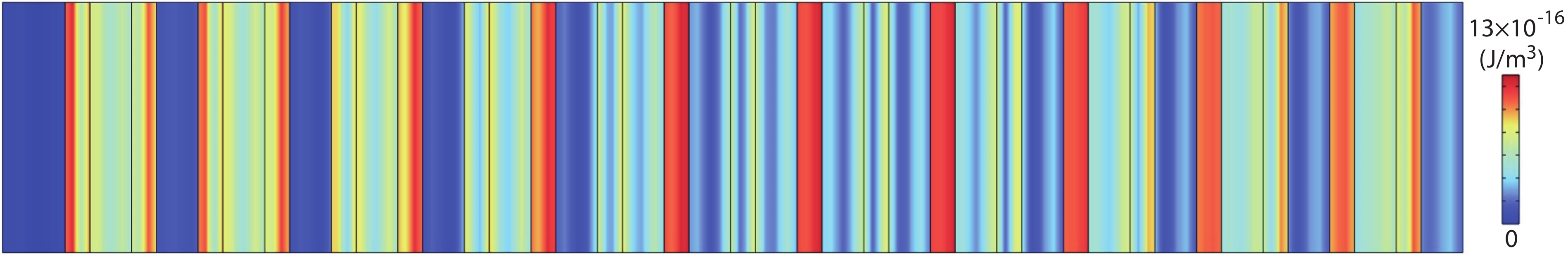}}
	\subfigure[]{\label{Fig:Res11e}
		\includegraphics[width=0.48\columnwidth]{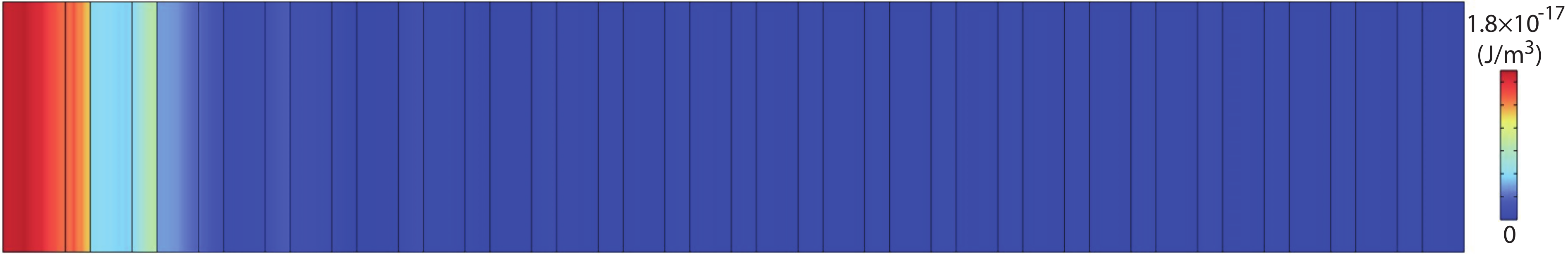}}
	\subfigure[]{\label{Fig:Res11f}
		\includegraphics[width=0.48\columnwidth]{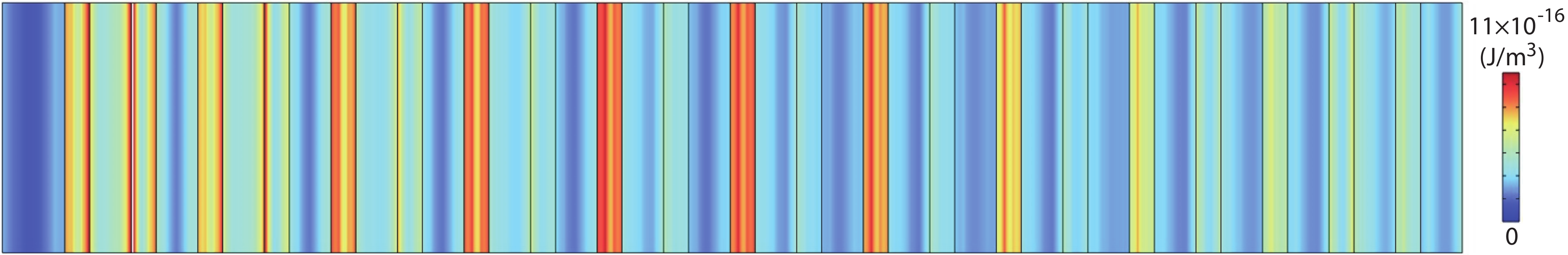}}
	\caption{Time-averaged power flow and energy distribution in the Bragg grating in Fig.~\ref{Fig:Res1}. (a)-(c)~Power flow at frequencies 150, 300 and 450 THz, respectively. (d)-(f)~Energy density at 150, 300 and 450 THz, respectively.\vspace{-0.5cm}}
	\label{Fig:Res11}
\end{center}
\end{figure}

Figure~\ref{Fig:Res11} presents the time-averaged power flow and electromagnetic energy density distribution within the coherent temporal Bragg grating from Figure~\ref{Fig:Res1}, analyzed at the three characteristic frequencies: the amplification band at $f_{\text{B}}/2$, the reflection band at $f_{\text{B}}$, and the second amplification band at $1.5f_{\text{B}}$. At 150 THz (Fig.~\ref{Fig:Res11a}), the power flow profile shows a significant net positive flux that grows along the propagation direction. This monotonic increase is a direct signature of distributed parametric amplification occurring throughout the grating, consistent with the field amplification seen in Fig.~\ref{Fig:Res1a}. Energy is being added to the wave via the temporal modulation of the low-index layers.
At 300~THz (Fig.~\ref{Fig:Res11b}), the power flow exhibits a characteristic standing-wave pattern with a vanishing net flux, where transmission is inhibited. At 450~THz (Fig.~\ref{Fig:Res11c}), similar to the 150 THz case, the power flow demonstrates a clear amplifying trend, increasing with propagation distance. This confirms that the gain mechanism is active not only at the sub-harmonic \((f_{\text{B}}/2)\) but also at a super-harmonic \((1.5f_{\text{B}})\) frequency which shows the frequency-symmetric nature of the parametric process driven by modulation at \(f_{\text{m}} = f_{\text{B}}\).

The electromagnetic energy density at 150 and 450 THz (Figs.~\ref{Fig:Res11d} and~\ref{Fig:Res11f}) shows a pronounced and growing concentration within the time-modulated low-index layers (corresponding to the regions with \(\delta_{\text{L}} > 0\)). This spatial localization and growth of energy directly correlate with the zones of active power injection, visualizing how the temporal modulation pumps energy into the system at these detuned frequencies. At 300 THz (Fig.~\ref{Fig:Res11e}), the energy density forms a stationary, high-contrast interference pattern typical of a Bragg reflector.

The results in Fig.~\ref{Fig:Res11} provide a complete spatial-energy picture of the coherent temporal grating's operation. They visually corroborate the transmission spectrum from Fig.~\ref{Fig:Res1d} as follows. i)~Amplification Bands: Net growing power flow and concurrent energy buildup in the modulated layers at frequencies detuned from \(f_{\text{B}}\), i.e., \(f_{\text{B}}/2\) and \(1.5f_{\text{B}}\). ii)~Photonic Bandgap: Resonant reflection with zero net power flow and a stationary energy pattern at the Bragg frequency (300 THz). This analysis confirms that a temporal modulation at the Bragg frequency \(f_{\text{m}} = f_{\text{B}}\) fundamentally transforms a static photonic crystal from a passive reflector into an active, distributed parametric amplifier for frequencies outside the original bandgap, while preserving perfect reflection at the central frequency.

\begin{figure}
	\begin{center}
			\includegraphics[width=0.6\columnwidth]{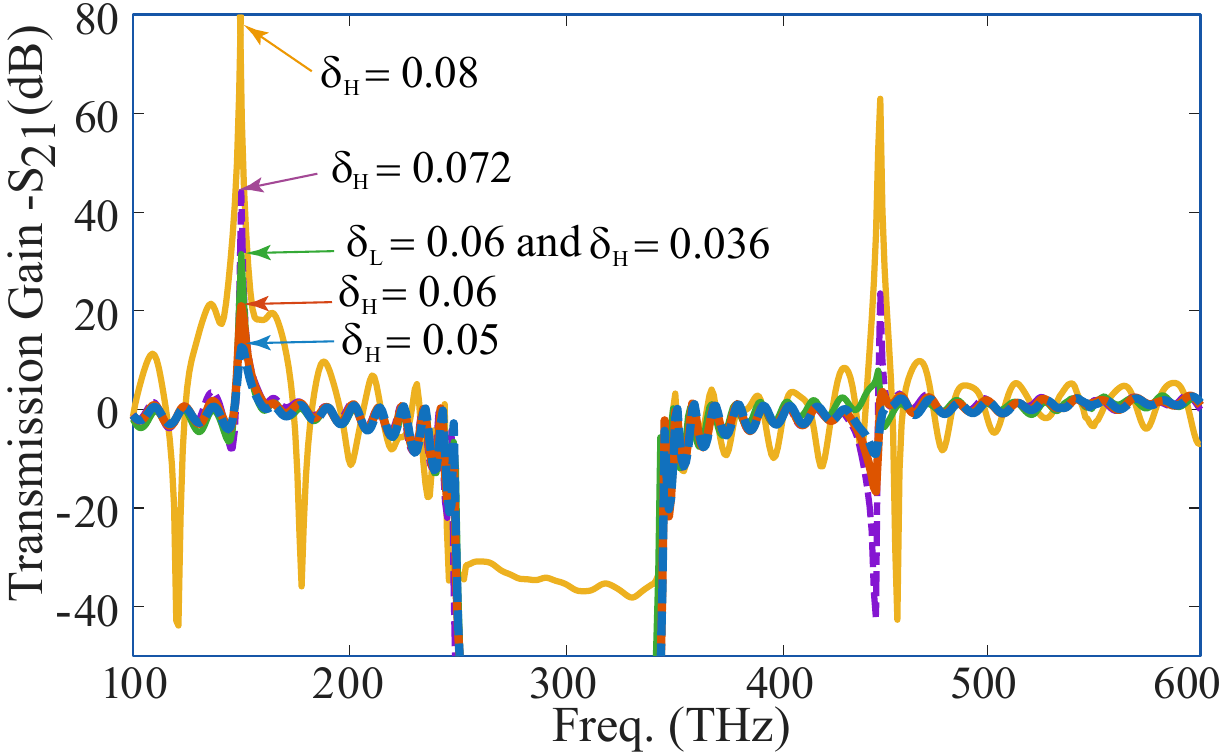}
\caption{Power amplification in a coherent temporal Bragg grating with modulation frequency \(f_\text{m} = f_\text{B}\). Spectra are shown for the case where the low-index layers are static (\(\delta_L = 0\)) and the high-index layers are time-modulated with varying amplitudes \(\delta_H\), along with one case of concurrent modulation (\(\delta_L = 0.06, \delta_H = 0.036\)). All other parameters match those in Fig. 2. The plot demonstrates strong parametric gain at $f_\text{B}/2$ and $1.5f_\text{B}$.\vspace{-0.5cm}}
		\label{Fig:Res2}
	\end{center}
\end{figure}

Figure~\ref{Fig:Res2} examines the impact of modulating the high-index layers on power amplification in a coherent temporal Bragg grating, under the condition \(f_{\text{m}} = f_{\text{B}}\). The parameters are the same as in Fig.~\ref{Fig:Res1}, but with a key modification: the low-index layers are now static (\(\delta_{\text{L}} = 0\)), while the high-index layers undergo time-periodic modulation with varying amplitudes \(\delta_{\text{H}} > 0\). An additional case with concurrent modulations (\(\delta_{\text{L}} = 0.06,\ \delta_{\text{H}} = 0.036\)) is also shown.

The spectra for high-index layer modulation (\(\delta_{\text{L}} = 0, \delta_{\text{H}} > 0\)) confirm that amplification persists at the same characteristic frequencies of $f_{\text{B}}/2$ and $1.5f_{\text{B}}$, demonstrating that the parametric gain mechanism is not exclusive to low-index modulation.
The gain magnitude increases significantly with larger modulation amplitudes \(\delta_{\text{H}}\). For instance, the gain at both $f_{\text{B}}/2$ and $1.5f_{\text{B}}$ for \(\delta_{\text{H}} = 0.08\) exceeds 60 dB, substantially higher than the gains observed in Fig.~\ref{Fig:Res2} for comparable \(\delta_{\text{L}}\) values. This indicates that modulating the high-index layers can be a more efficient method for generating parametric amplification in this structure.
The reflection bandgap at the central Bragg frequency remains intact, maintaining near-zero transmission. The curve for concurrent modulation case (\(\delta_{\text{L}} = 0.06, \delta_{\text{H}} = 0.036\)) in Fig.~\ref{Fig:Res2} shows a remarkable gain profile for such small modulation strengths. The achieved amplification is lower than for the largest pure \(\delta_{\text{H}}=0.08\) modulations but demonstrates that the effect is the same when both layers are weakly modulated.
This suggests that the total modulation strength and its distribution influence the final gain. The gain does not simply result from adding the effects of individual modulations; it depends on their relative phases and amplitudes within the grating's temporal dynamics. The results in Fig.~\ref{Fig:Res2} reveal that time-varying high-index layers are highly effective for generating parametric amplification in a coherent temporal Bragg grating. The gain mechanism remains frequency-symmetric around \(f_{\text{B}}\), producing strong amplification at \(f_{\text{B}}/2\) and \(1.5f_{\text{B}}\).

\begin{figure}
	\begin{center}
		\valign{#\cr
			\hsize=0.5\columnwidth
			\subfigure[]{\label{Fig:Res3a}
				\includegraphics[width=.26\textwidth,height=3.2cm]{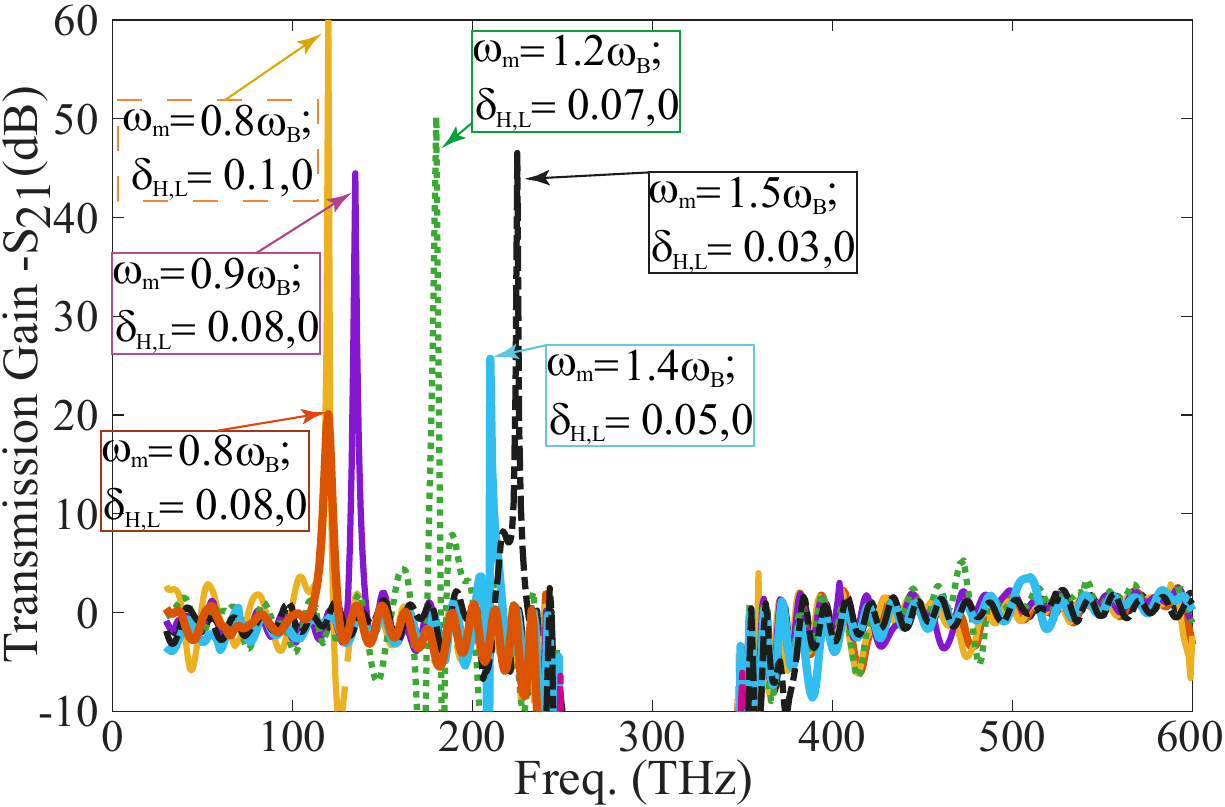}}\cr\noalign{\hfill}
			\hsize=0.45\columnwidth
			\subfigure[]{\label{Fig:Res3b}
				\includegraphics[width=0.425\columnwidth]{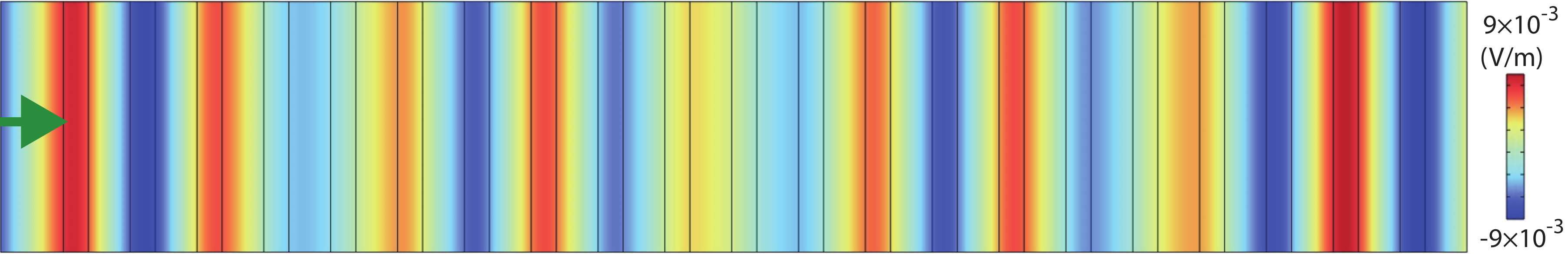}}
			\subfigure[]{\label{Fig:Res3c}
				\includegraphics[width=0.425\columnwidth]{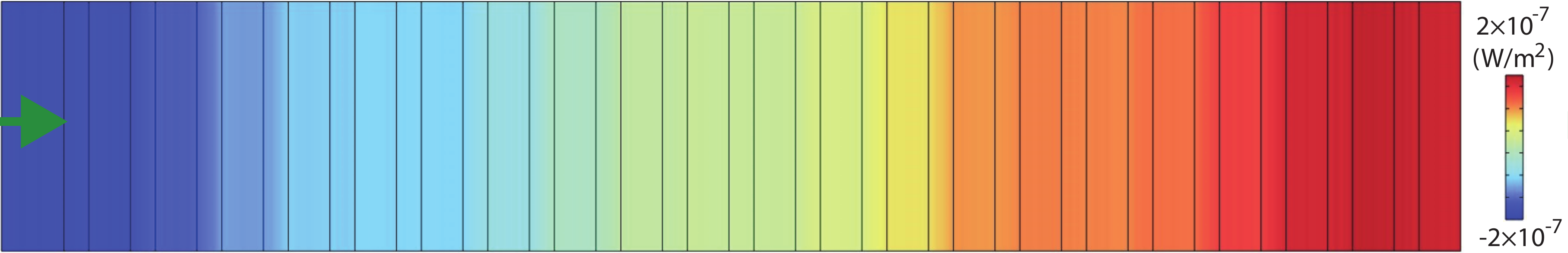}}
		\subfigure[]{\label{Fig:Res3d}
			\includegraphics[width=0.425\columnwidth]{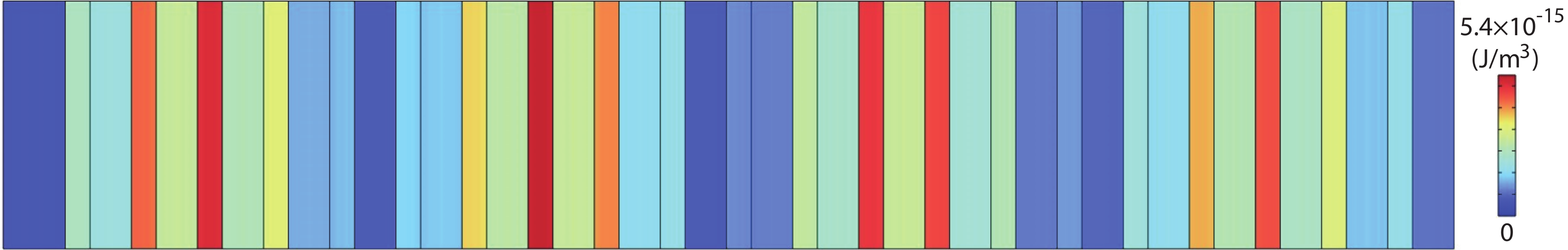}}\cr}\vspace{-0.5cm}
		\caption{Frequency-agile power amplification in a temporal Bragg grating with the same parameters as in Fig.~\ref{Fig:Res1}, except for distinct modulation frequencies of $f_\text{m}=f_\text{B}\times[0.8,0.9,1.2,1.4,1.5]$ corresponding to distinct peak amplification frequencies of $f=[120,135,180,210,225]$~THz. (a)~Transmission gain spectra. (b)~Normal electric field, (c)~Time-averaged power flow, and (d)~Electromagnetic energy density.\vspace{-0.5cm}}
		\label{Fig:Res3}
	\end{center}
\end{figure}

Figure~\ref{Fig:Res3} demonstrates frequency-agile parametric amplification in a temporal Bragg grating, achieved by decoupling the modulation frequency \(f_{\text{m}}\) from the Bragg frequency \(f_{\text{B}}\). Here, the system parameters remain the same as in Fig.~\ref{Fig:Res1}, except that the modulation frequency is varied relative to \(f_{\text{B}}\). The results show that the amplification peaks are not fixed at \(f_{\text{B}}/2\) and \(1.5f_{\text{B}}\), but instead shift dynamically with \(f_{\text{m}}\), enabling tunable gain across a broad spectral range. The figure clearly illustrates that the amplification frequency is directly controlled by the modulation frequency \(f_{\text{m}}\). Each curve corresponds to a different modulation frequency, expressed as a fraction or multiple of \(f_{\text{B}}\), where for \(f_{\text{m}} = 0.8f_{\text{B}}\), the peak amplification occurs at \(f = 120\ \text{THz}\), for \(f_{\text{m}} = 0.9f_{\text{B}}\), the peak shifts to \(f = 135\ \text{THz}\), for \(f_{\text{m}} = 1.2f_{\text{B}}\), amplification is observed at \(f = 180\ \text{THz}\), for \(f_{\text{m}} = 1.4f_{\text{B}}\), the peak moves to \(f = 210\ \text{THz}\), and for \(f_{\text{m}} = 1.5f_{\text{B}}\), amplification occurs at \(f = 225\ \text{THz}\).
Figure~\ref{Fig:Res3} confirms parametric amplification via phase-matching: gain peaks at \(f_{\text{signal}} = f_{\text{B}} \pm f_{\text{m}}\), corresponding to Stokes and anti-Stokes sidebands generated by temporal modulation. Key implications are i)~Tunability, where adjusting \(f_{\text{m}}\) continuously tunes the amplification band, enabling reconfigurable amplifiers and frequency converters. ii)~Beyond coherent modulation (\(f_{\text{m}} = f_{\text{B}}\)), detuning breaks symmetry, placing gain asymmetrically relative to the bandgap. iii)~The original photonic bandgap near \(f_{\text{B}}=300\) THz remains intact—temporal modulation primarily affects frequencies outside the static gap which enables simultaneous reflection at \(f_{\text{B}}\) and amplification elsewhere.

\subsection{Regime II: Sub-Bragg Modulation}

\begin{figure}
	\begin{center}
		\valign{#\cr
			\hsize=0.5\columnwidth
			\subfigure[]{\label{Fig:Res4a}
				\includegraphics[width=.26\textwidth,height=3.2cm]{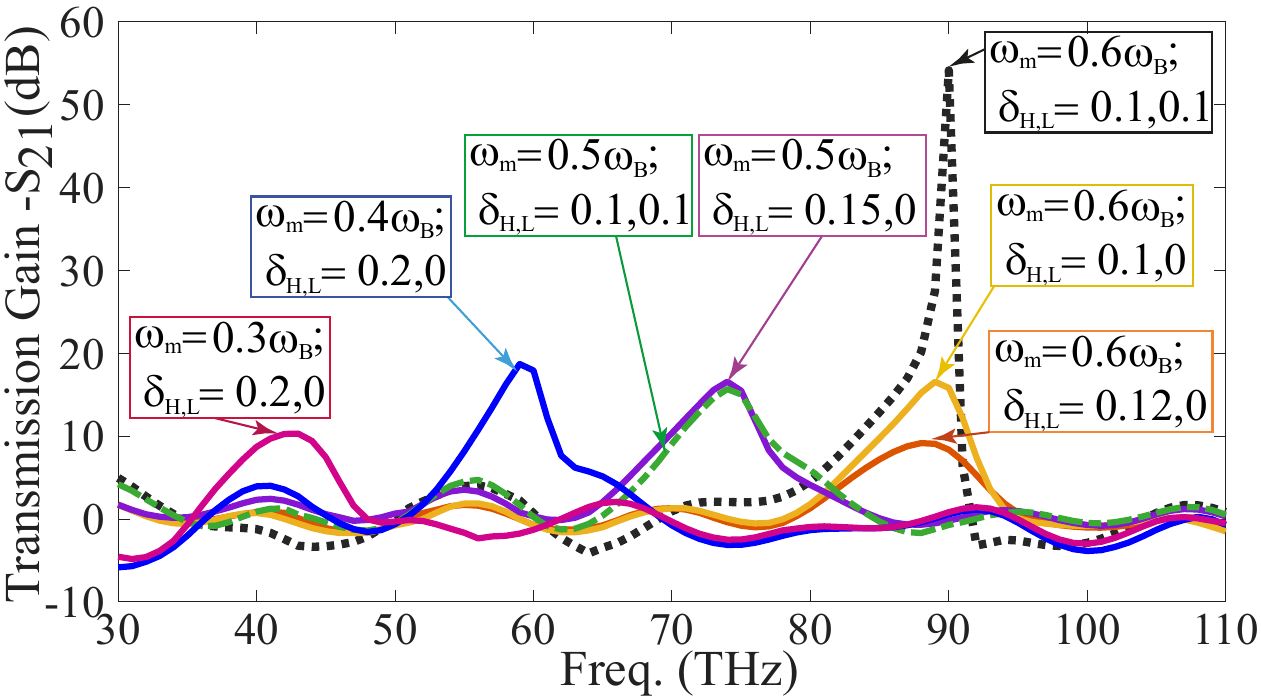}}\cr\noalign{\hfill}
			\hsize=0.45\columnwidth
			\subfigure[]{\label{Fig:Res4b}
				\includegraphics[width=0.425\columnwidth]{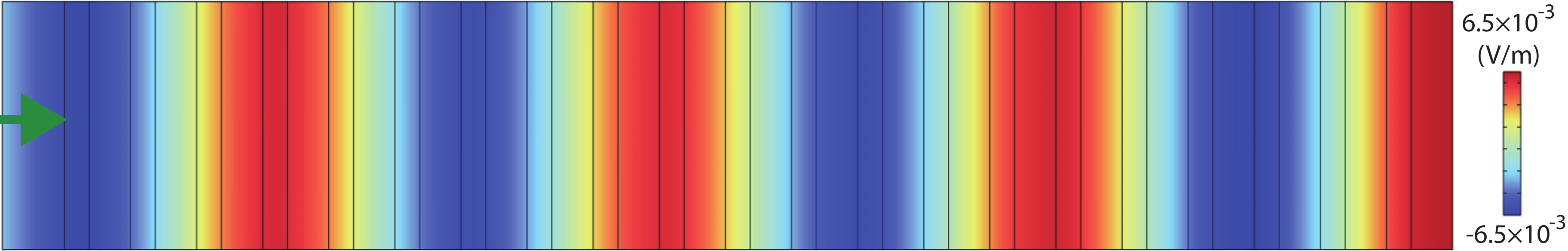}}
			\subfigure[]{\label{Fig:Res4c}
				\includegraphics[width=0.425\columnwidth]{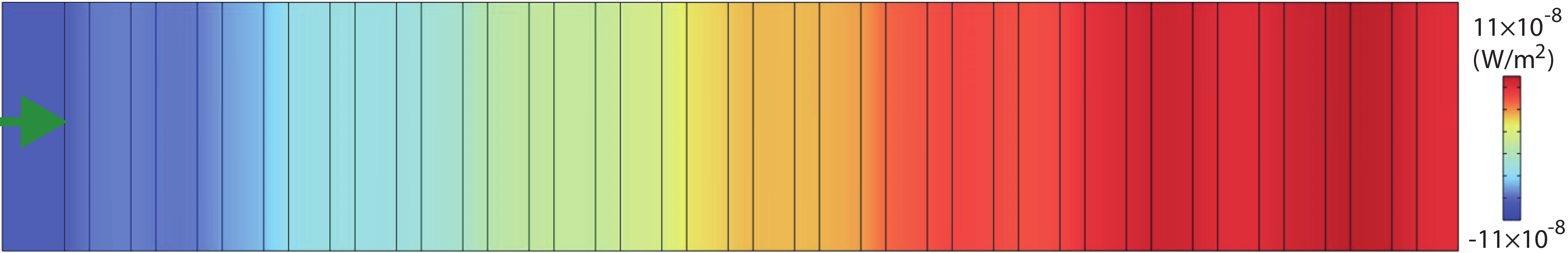}}
			\subfigure[]{\label{Fig:Res4d}
				\includegraphics[width=0.425\columnwidth]{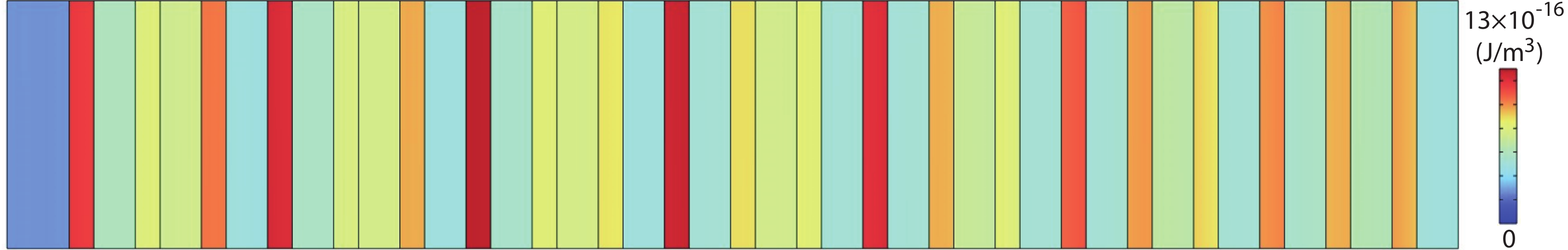}}\cr}\vspace{-0.5cm}
		\caption{Power amplification in the sub-Bragg modulation regime (\(f_{\text{m}} < f_{\text{B}}\)). (a)~Transmission gain spectra for modulation frequencies \(f_{\text{m}} = f_{\text{B}} \times [0.3, 0.4, 0.5, 0.6]\), showing corresponding amplification peaks at \(f = [45, 60, 75, 90]\) THz. Significant gain in this regime requires strong modulation amplitudes (\(\delta_{\text{L}} = \delta_{\text{H}} = 0.1\)). (b)-(d) Spatial distributions for the case \(f_{\text{m}} = 0.6f_{\text{B}}\) at its peak frequency \(f = 90\) THz: (b)~Normal electric field, (c)~Time-averaged power flow, and (d)~Electromagnetic energy density.\vspace{-0.5cm}}
		\label{Fig:Res4}
	\end{center}
\end{figure}

Figure~\ref{Fig:Res4} explores the regime of sub-Bragg modulation, where the modulation frequency is lower than the Bragg frequency (\(f_\text{m} < f_\text{B}\)). The study reveals that achieving significant parametric amplification in this region requires substantially stronger modulation strengths compared to the supra-Bragg case (\(f_\text{m} > f_\text{B}\)) which highlights an asymmetry in the system's nonlinear response. The transmission gain spectra in Fig.~\ref{Fig:Res4a} show clear amplification peaks at frequencies directly tied to the modulation frequency: \(f = 45, 60, 75,\) and \(90\)~THz for \(f_\text{m} = 0.3f_\text{B}, 0.4f_\text{B}, 0.5f_\text{B},\) and \(0.6f_\text{B}\), respectively. This confirms the frequency-agile nature of the gain. To achieve gain levels on the order of tens of dB in this sub-Bragg regime, the modulation amplitudes must be set significantly higher (\(\delta_\text{L} = \delta_\text{H} = 0.1\)) than those typically sufficient for the coherent (\(f_\text{m}=f_\text{B}\)) or supra-Bragg cases shown in previous figures. This indicates a reduced parametric coupling efficiency when \(f_\text{m} < f_\text{B}\).

The electric field and energy dynamics at \(f = 90\)~THz in Figs.~\ref{Fig:Res4b}-\ref{Fig:Res4d}) show that for the case \(f_\text{m} = 0.6f_\text{B}\) and strong modulation (\(\delta_\text{L} = \delta_\text{H} = 0.1\)), the spatial distributions at the amplification frequency (\(90\) THz) provide insight into the operative physics. The normal electric field distribution in Fig.~\ref{Fig:Res4b} shows a propagating wave with a growing envelope, characteristic of distributed amplification. The pattern confirms that efficient power transfer is occurring despite the sub-Bragg modulation condition. The time-averaged power flow in Fig.~\ref{Fig:Res4c} exhibits a clear monotonic increase along the propagation direction. This positive, growing flux is the definitive signature of net power amplification supplied by the temporal modulation. The energy density in Fig.~\ref{Fig:Res4d} confirms that the electromagnetic energy density is strongly localized and enhanced within the modulated layers of the grating.

The need for stronger modulation in the sub-Bragg regime stems from phase-matching and the photonic density of states near the bandgap. For \(f_\text{m} < f_\text{B}\), sidebands lie closer to the gap center, where reduced group velocity and distorted dispersion weaken coupling which requires larger modulation depth to achieve comparable gain. Figure~\ref{Fig:Res4} confirms that while frequency-agile amplification is possible for \(f_\text{m} < f_\text{B}\), it demands significantly stronger modulation which reveals an inherent asymmetry between sub-Bragg and supra-Bragg regimes governed by the static crystal's dispersion. This dictates distinct pump power requirements depending on the desired operational band relative to \(f_\text{B}\), a critical design insight for practical tunable amplifiers.

\subsection{Regime III: Extreme Sub-Bragg Modulation}

Figure~\ref{Fig:Res5} investigates the extreme sub-Bragg modulation regime, where the modulation frequency is a small fraction of the Bragg frequency (\(f_{\text{m}} = 0.1f_{\text{B}}\) and \(0.2f_{\text{B}}\)). This regime reveals a significant transition in the system's behavior: while discrete, frequency-specific amplification peaks persist, a new broadband gain continuum emerges at high frequencies (\(\gtrsim f_{\text{B}}\)), particularly for the smallest modulation frequency (\(f_{\text{m}} = 0.1f_{\text{B}}\)).

\textit{Persistence of Discrete Amplification with Strong Modulation:} As shown in Fig.~\ref{Fig:Res5a}, discrete amplification peaks are still generated at frequencies directly linked to the modulation (e.g., \(f \approx 12\) THz for \(f_{\text{m}} = 0.2f_{\text{B}}\) and \(f \approx 559\) THz for \(f_{\text{m}} = 0.1f_{\text{B}}\)). Achieving substantial gain at these discrete peaks requires exceptionally strong modulation amplitudes (\(\delta_{\text{H}} = 0.2\) to \(0.4\)), consistent with the reduced parametric coupling efficiency in the deep sub-Bragg regime.

The spatial distributions confirm the amplification physics for these discrete peaks:
For the 12 THz peak (\(f_{\text{m}}=0.2f_{\text{B}}\), Figs.~\ref{Fig:Res5b}-\ref{Fig:Res5d}): The electric field (b) shows growth, the power flow (c) increases, and the energy density (d) localizes, confirming parametric gain.
For the 559 THz peak (\(f_{\text{m}}=0.1f_{\text{B}}\), Figs.~\ref{Fig:Res5e}-\ref{Fig:Res5g}): Similar characteristics of field growth (e), positive power flux (f), and localized energy (g) are observed.

\textit{Multi-Phase-Matching as the Origin of the Broadband High-Frequency Gain}: The most significant feature in Fig.~\ref{Fig:Res5a} is the appearance of a broad, continuous band of gain at high frequencies. This broadband effect is not due to a single parametric process but results from multiple non-degenerate four-wave mixing processes that become simultaneously phase-matched when the modulation frequency is very small.

Physically, for a small modulation wavenumber \(K = \omega_{\text{m}} n_g / c\), the phase-matching condition for parametric amplification can be approximated as
\begin{equation}
	\beta(\omega_0 + \Delta\omega) + \beta(\omega_0 - \Delta\omega) \approx 2\beta_0 + K,
\end{equation}
where \(\beta_0 = \beta(\omega_0)\) is the propagation constant at the Bragg frequency. For a sufficiently small \(K\), this condition can be satisfied for a continuous range of frequency detunings \(\Delta\omega\) within the system's phase-matching bandwidth \(\Delta\omega_{\text{PM}} \approx \pi c/(n_g L)\), where \(L\) is the grating length and \(n_g\) the group index. Thus, as \(f_{\text{m}} \to 0\), the modulation wavevector \(K\) becomes negligible compared to the dispersion scale. This allows a spectrum of signal-idler pairs around \(f_0\) to simultaneously satisfy the phase-matching condition, resulting in the observed broadband gain continuum rather than discrete sidebands.

\section{Discussions}
\begin{figure}
	\begin{center}
		\subfigure[]{\label{Fig:Res5a}
			\includegraphics[width=0.7\columnwidth]{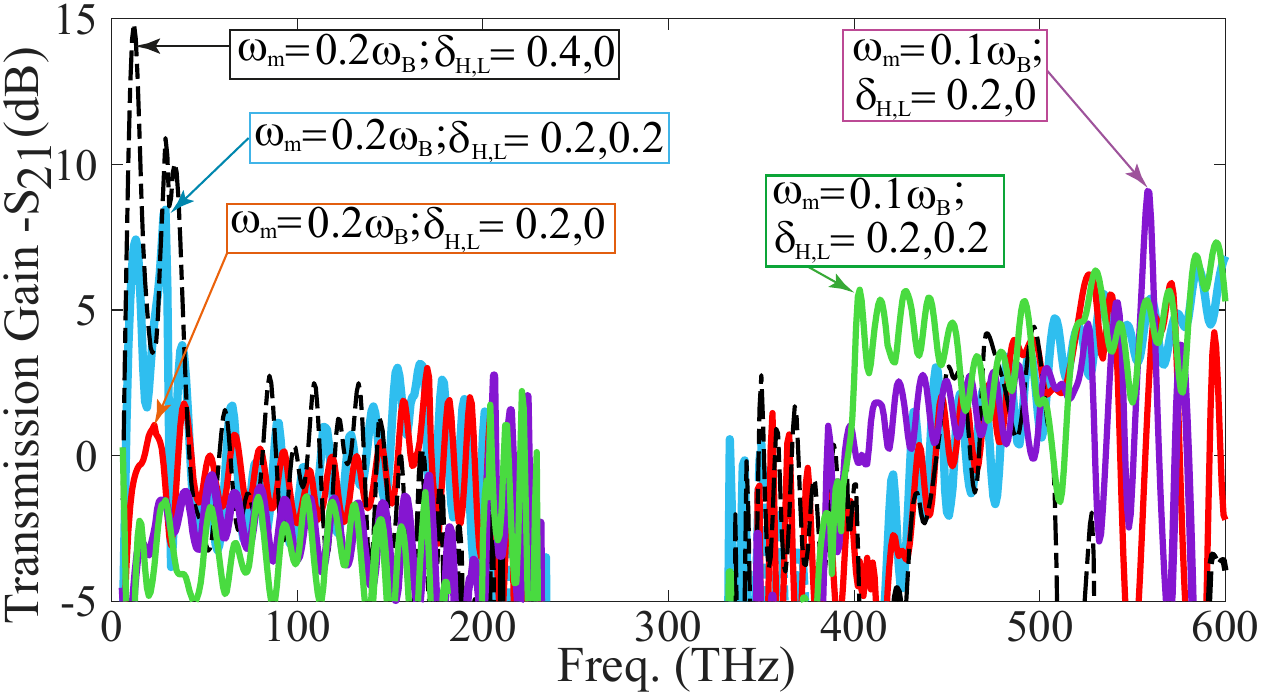}}
   		\subfigure[]{\label{Fig:Res5b}
   	\includegraphics[width=0.48\columnwidth]{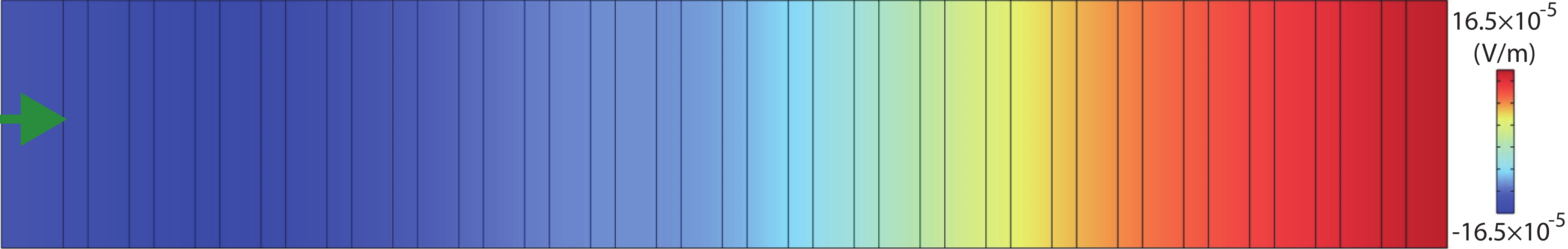}}
   \subfigure[]{\label{Fig:Res5c}
   	\includegraphics[width=0.48\columnwidth]{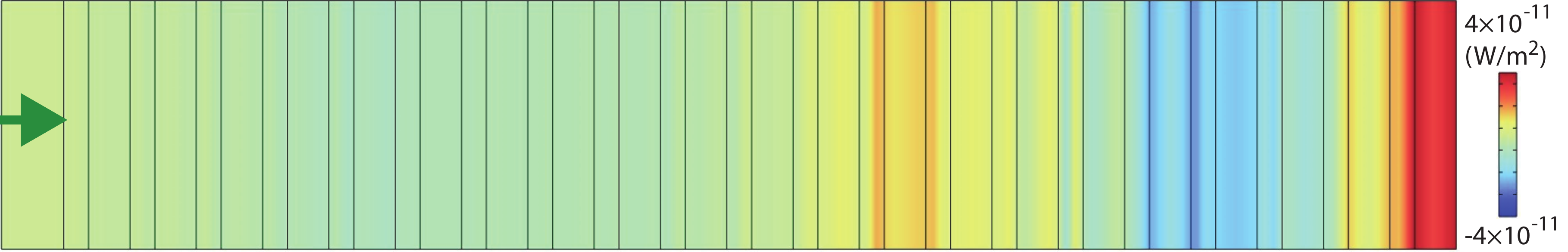}}
   \subfigure[]{\label{Fig:Res5d}
   	\includegraphics[width=0.48\columnwidth]{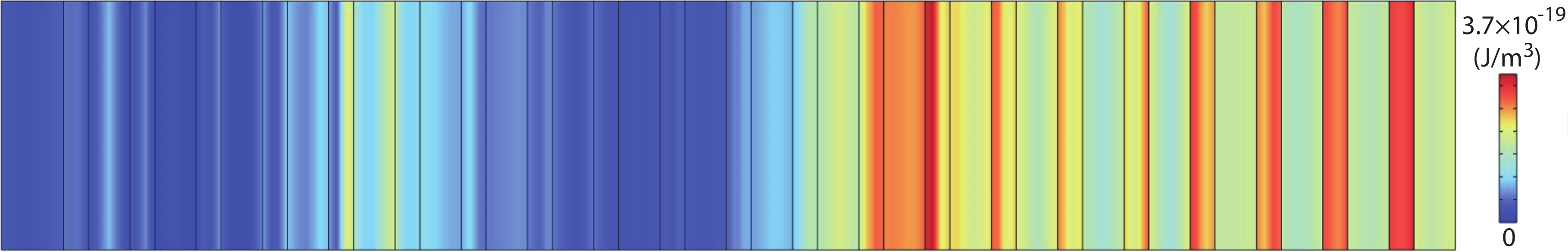}}
		\subfigure[]{\label{Fig:Res5e}
		\includegraphics[width=0.48\columnwidth]{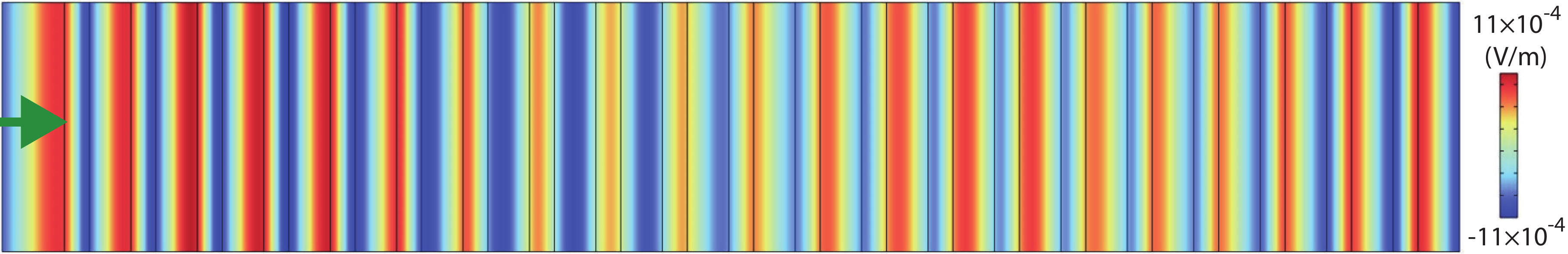}}
		\subfigure[]{\label{Fig:Res5f}
		\includegraphics[width=0.48\columnwidth]{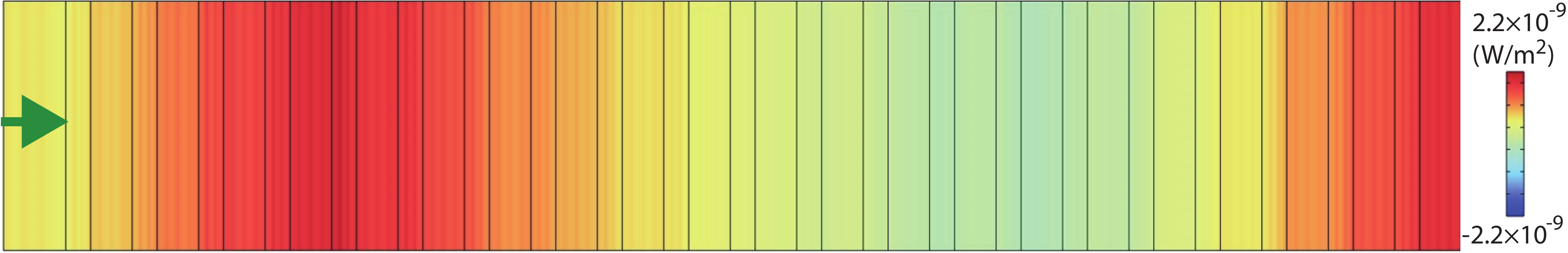}}
		\subfigure[]{\label{Fig:Res5g}
		\includegraphics[width=0.48\columnwidth]{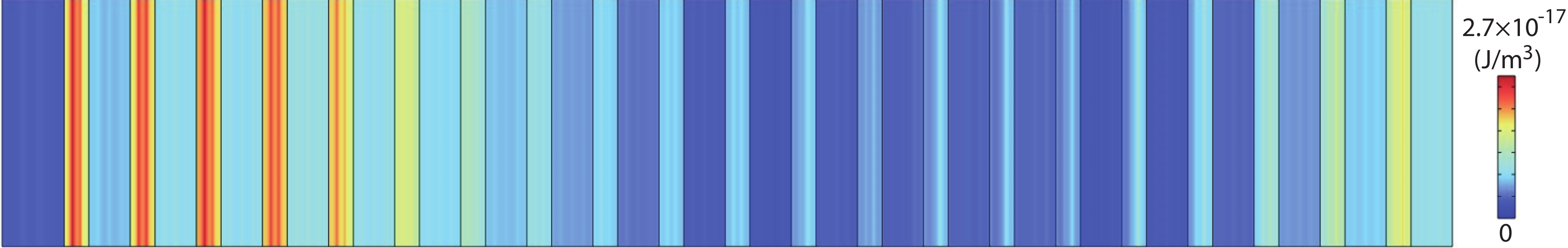}}
		\caption{Frequency-agile power amplification in a temporal Bragg grating with the same parameters as in Fig.~\ref{Fig:Res1}, except for varying modulation amplitudes at distinct modulation frequencies of $f_\text{m}=f_\text{B}\times[0.1,0.2]$ corresponding to distinct peak amplification frequencies of $f=[12,559]$~THz. (a)~Transmission gain. (b)-(d)~Normal electric field distribution, time-average power flow, and energy density at $f=12$~THz for the grating with $f_\text{m}=0.2f_\text{B}$, $\delta_\text{H}=0.4$ and $\delta_\text{L}=0$. (e)-(g)~Normal electric field distribution, time-average power flow, and energy density at $f=559$~THz for the grating with $f_\text{m}=0.1f_\text{B}$, $\delta_\text{H}=0.2$ and $\delta_\text{L}=0$.\vspace{-0.5cm}}
		\label{Fig:Res5}
	\end{center}
\end{figure}

Figure~\ref{Fig:Res5} reveals a fundamental operational transition governed by modulation frequency. For \(f_{\text{m}} \gtrsim 0.2f_{\text{B}}\), the system acts as a discrete, frequency-agile parametric amplifier requiring stronger modulation as \(f_{\text{m}}\) decreases. For \(f_{\text{m}} \lesssim 0.1f_{\text{B}}\), multi-phase-matching enables a broadband amplifier regime where a single low-frequency pump provides gain across a wide high-frequency band. Thus, modulation frequency directly controls gain bandwidth: engineers can select between narrowband tunable amplification (\(f_{\text{m}} \sim f_{\text{B}}\)) and wideband fixed-band amplification (\(f_{\text{m}} \ll f_{\text{B}}\)) by adjusting a single parameter.

The modulation depths employed in this study ($\delta < 0.12$ for $<20$ dB amplification) are achievable through several established techniques. Electro-optic materials (e.g., LiNbO$_3$) provide $\delta\sim 0.01–0.1$ at GHz frequencies~\cite{wooten2000review,wang2018integrated}, while carrier injection in silicon photonics yields $\delta\sim0.1–0.5$~\cite{soref1987electrooptical,reed2010silicon}. All-optical approaches achieve $\delta\sim0.01–0.1$ at THz speeds \cite{foster2008silicon}. Recent experiments on electrically tunable space-time metasurfaces at $\lambda = 1530$ nm have demonstrated MHz modulation with ITO-based plasmonic structures, achieving effective index changes exceeding $\delta \sim 0.2$ and confirming the feasibility of spatiotemporal control at optical frequencies~\cite{amin2020sub,shaltout2019spatiotemporal,huang2016gate,sisler2024electrically}. Modulation frequencies up to $0.3\omega_0$ (tens of THz) are accessible via nonlinear waveguides~\cite{nahata1996wideband}.

In the weak-coupling regime where $\kappa d \ll 1$ (with $\kappa \propto \delta$ the coupling coefficient and $d$ the layer thickness), each period contributes a small increment to the amplified wave. The total field after $N$ periods is the coherent sum of these contributions, leading to \textbf{linear scaling} with $N$, as $A_{\text{amp}} \propto N \cdot \delta \cdot A_0$, where $A_0$ is the incident amplitude. This linear regime corresponds to the perturbative limit where pump depletion and back-conversion are negligible. As the modulation depth or number of periods increases, the system enters the exponential gain regime. This transition occurs when the parametric coupling becomes strong enough that the amplified field significantly depletes the incident wave, and multiple scattering events between harmonics must be considered. Solving the coupled-mode equations without the undepleted-pump approximation yields
\begin{equation}
	A_{\text{amp}}(z) = A_0 \cosh(g z) \approx \frac{A_0}{2} e^{g z} \quad \text{for} \quad g N \Lambda \gg 1,
	\label{eq:exponential_scaling}
\end{equation}
where $g = |\kappa| \cdot \eta$ is the parametric gain coefficient, with $\eta$ representing the spatial overlap factor between the interacting modes. The transition between regimes occurs when the gain-length product satisfies $g L \sim 1$, corresponding to
\begin{equation}
	N_{\text{crit}} \approx \frac{1}{g \Lambda} \propto \frac{1}{\delta},
	\label{eq:critical_N}
\end{equation}
where for $N \ll N_{\text{crit}}$, the system operates in the linear regime; for $N \gg N_{\text{crit}}$, exponential growth dominates. In the linear regime, gain increases slowly with $N$ and requires many periods to achieve modest amplification. However, once the critical threshold is crossed, dramatic amplification becomes possible with relatively few additional periods. The modulation depth $\delta$ directly controls the critical number of periods $N_{\text{crit}}$, providing a design knob to tailor the device's operating regime.

The bandwidth of the broadband gain region can be further expanded through grating optimization. Strategies include: (i) apodized gratings to flatten the gain profile; (ii) chirped gratings with graded layer thicknesses to extend phase-matching, analogous to broadband quasi-phase-matching; (iii) dispersion engineering to reduce group velocity dispersion near the band edge; (iv) shorter gratings to broaden bandwidth at the expense of peak gain; and (v) cascaded modulation profiles to synthesize ultra-broadband responses. These strategies, informed by multi-phase-matching, provide a roadmap for tailoring gain bandwidth to specific applications.

To contextualize our work, we compare temporal Bragg grating amplifiers with conventional technologies. Fiber amplifiers offer high gain and low noise but require meter-scale lengths and lack tunability. Semiconductor optical amplifiers are compact but exhibit higher noise and limited integration due to III-V materials. Temporal Bragg gratings uniquely combine ultra-compact footprint ($<10~\mu$m), outstanding gain density of $\sim$3.7 dB/$\mu$m, and broad tunability via modulation frequency $\omega_\text{m}$. The parametric mechanism ensures quantum-limited noise performance, while dynamic reconfigurability between narrowband and broadband operation, impossible in conventional fixed-gain media, enables frequency-agile amplification. With CMOS-compatible materials and moderate modulation depths ($\delta<0.12$), this platform is promising for integrated RF photonic links and signal processing.

While our analysis assumes lossless media for clarity, the formalism readily extends to lossy media via complex propagation constants $\tilde{\beta}_{i,q} = \beta_{i,q} + i\alpha_i/2$. For typical silicon photonics losses ($\alpha\sim1$–$3$ dB/cm) and mm-scale devices, the gain reduction is negligible ($<0.1$ dB), and the threshold modulation depth ($\delta_{\text{th}} \sim 10^{-5}$) lies far below our operating $\delta\sim0.12$.

\vspace{-0.2cm}
\section{Conclusion}\label{sec:conc}
We have introduced temporal Bragg gratings and demonstrated their capability for efficient parametric power amplification, transforming a static photonic crystal from a passive filter into an active, reconfigurable amplifier. Key findings include: (i) coherent amplification at \(f_\text{m}=f_\text{B}\) yields strong gain at \(f_\text{B}/2\) and \(1.5f_\text{B}\); (ii) spectral agility via \(f_\text{m}\) enables tunable gain at \(f_\text{B}\pm f_\text{m}\); (iii) sub-Bragg (\(f_\text{m}<f_\text{B}\)) operation requires stronger modulation than supra-Bragg; and (iv) at \(f_\text{m}\lesssim0.1f_\text{B}\), multi-phase-matching produces a broadband gain continuum. Temporal Bragg gratings thus offer a versatile platform dynamically reconfigurable between narrowband and broadband amplification via modulation frequency control—opening avenues for integrated multifunctional photonic devices and establishing time as a fundamental dimension in photonic engineering.

%
%

%
\bibliographystyle{IEEEtran}
\bibliography{Taravati_Reference.bib}

\vspace{-1cm}
\begin{IEEEbiography}[{\includegraphics[width=0.95in,height=1.2in,clip,keepaspectratio]{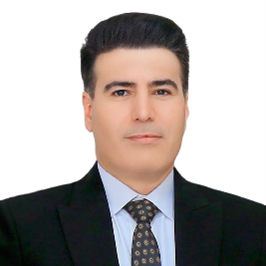}}]{Sajjad Taravati} (Senior Member, IEEE) is an Assistant Professor (UK Lecturer) at the University of Southampton, UK, leading a research group on dynamic metasurfaces for next-generation quantum and communication systems. Over the past decade, his work at the University of Toronto, the University of Montréal, Concordia University, the University of Oxford, and the University of Southampton has advanced nonreciprocal electromagnetics, spatiotemporal metasurfaces, and quantum computing, yielding over 90 publications and patents. He is the Technological Founder of LATYS Intelligence, commercializing dynamic metasurface technologies based on his patents. 
\end{IEEEbiography}

\end{document}